\documentclass{aa}

\usepackage{txfonts}
\usepackage{natbib}
\usepackage{units}
\usepackage{graphicx}
\usepackage{url}
\bibpunct{(}{)}{;}{a}{}{,}

\newcommand{\bea}{\begin{eqnarray*}}
\newcommand{\eea}{\end{eqnarray*}}
\newcommand{\bean}{\begin{eqnarray}}
\newcommand{\eean}{\end{eqnarray}}
\newcommand{\Div}{\vec{\nabla}\cdot}
\newcommand{\Grad}{\vec{\nabla}}
\newcommand{\Rot}{\vec{\nabla}\times}
\newcommand{\dH}{\mathbf{\nabla_H}}
\newcommand{\delh}{\mathbf{\nabla_h}}
\newcommand{\dd}[2]{\frac{\partial #1}{\partial #2}}
\newcommand{\ddt}[1]{\frac{\partial #1}{\partial t}}
\newcommand{\ddx}[1]{\frac{\partial #1}{\partial \vec{x}}}
\newcommand{\ddxk}[1]{\dd{#1}{x_k}}
\newcommand{\ddz}[1]{\dd{#1}{z}}
\newcommand{\BB}{\vec{B}}
\newcommand\AM{{\mathbf{A}}}
\newcommand\CC{{\mathbf{C}}}
\newcommand\DD{{\mathbf{D}}}
\newcommand\FF{{\mathbf{F}}}
\newcommand\SM{{\mathbf{S}}}
\newcommand\UU{{\mathbf{U}}}

\newcommand{\uu}{\mathbf{u}}
\newcommand{\uH}{\mathbf{u_H}}
\newcommand{\BH}{\mathbf{B_H}}
\newcommand{\Bh}{\mathbf{B_h}}
\newcommand{\Rh}{\mathbf{R_h}}
\newcommand{\signBz}{\mathrm{sign}(B_z)}
\newcommand{\dPde}{\left(\dd{P}{e}\right)_\rho}
\newcommand{\dPdr}{\left(\dd{P}{\rho}\right)_e}
\newcommand{\om}[1]{\frac{#1}{\mu_0 \rho}}
\newcommand{\som}[1]{\frac{#1}{\sqrt{\mu_0 \rho}}}
\newcommand{\sm}{\sqrt{\mu_0 \rho}}
\newcommand{\half}{\frac{1}{2}}
\newcommand{\Eq}[1]{Eq.\ \ref{#1}}
\newcommand{\Eqs}[2]{Eq.\ \ref{#1}--\ref{#2}}
\newcommand{\Fig}[1]{Fig.\ \ref{#1}}

\newcommand{\Figs}[2]{Figs.\ \ref{#1}--\ref{#2}}
\newcommand{\Sec}[1]{Sect.\ \ref{#1}}
\newcommand{\no}{\ensuremath{n_\mathrm{o}}}
\newcommand{\nel}{\ensuremath{n_\mathrm{e}}}
\newcommand{\nhtwo}{\ensuremath{n_\mathrm{H2}}}
\newcommand{\none}{n_{1}}
\newcommand{\noneo}{n_{1}^\mathrm{o}}
\newcommand{\nhtwoo}{\ensuremath{n_\mathrm{H2}^\mathrm{o}}}
\newcommand{\htwo}{\ensuremath{\mathrm{H}_2}}
\newcommand{\rml}{\ensuremath{\mathrm{l}}}
\newcommand{\Cthreeh}{\ensuremath{C_\mathrm{3H}}}
\newcommand{\Chtwoh}{\ensuremath{C_\mathrm{H2H}}}
\def\specchar#1{\uppercase{#1}}    %% to be redefined for A&A, small caps
\newcommand\CaII{\mbox{Ca\,\specchar{ii}}} 
\newcommand\MgII{\mbox{Mg\,\specchar{ii}}} 
\newcommand\eg{e.g.,}              %% , required (Webster 1681)
\newcommand{\bifrost}{{\textsl{Bifrost}}}

\DeclareRobustCommand\sfrac[1]{\@ifnextchar/{\@sfrac{#1}}%
                                            {\@sfrac{#1}/}}
\def\@sfrac#1/#2{\leavevmode\kern.1em\raise.5ex
         \hbox{$\m@th\mbox{\fontsize\sf@size\z@
                           \selectfont#1}$}\kern-.1em
         /\kern-.15em\lower.25ex
          \hbox{$\m@th\mbox{\fontsize\sf@size\z@
                            \selectfont#2}$}}

\title{The stellar atmosphere simulation code \bifrost{}}

\subtitle{Code description and validation}
\author{B. V. Gudiksen\inst{1,2}
\and M. Carlsson\inst{1,2}
\and V. H. Hansteen\inst{1,2}
\and W. Hayek\inst{3}
\and J. Leenaarts\inst{4}
\and J. Mart\'inez-Sykora\inst{1,5}
}

\institute{
Institute of Theoretical Astrophysics, University of Oslo, P.O. Box
1029 Blindern, N-0315 Oslo, Norway
\and
Center of Mathematics for Applications, University of Oslo, P.O. Box
1053, Blindern, N-0316 Oslo, Norway
\and
School of Physics, University of Exeter, Stocker Road, Exeter EX4 4QL, United Kingdom
\and
Sterrekundig Instituut, Utrecht University, Postbus 80\,000,
NL--3508 TA Utrecht, The Netherlands
\and
Lockheed Martin Solar \& Astrophysics Lab, Org.\ ADBS,
Bldg.\ 252, 3251 Hanover Street Palo Alto, CA~94304 USA
}

\date{\today / -  }

\authorrunning{Gudiksen et al.}
\titlerunning{The numerical code \bifrost{}}
\abstract
{Numerical simulations of stellar convection and photospheres have been developed to the point where detailed shapes of observed spectral lines can be explained. Stellar atmospheres are very complex, and very different physical regimes are present in the convection zone, photosphere, chromosphere, transition region and corona. To understand the details of the atmosphere it is necessary to simulate the whole atmosphere since the different layers interact strongly. These physical regimes are very diverse and it takes a highly efficient massively parallel numerical code to solve the associated equations.} %Such simulations would demand extreme computing power and have so far not been published}
{The design, implementation and validation of the massively parallel numerical code \bifrost{} for simulating stellar atmospheres from the convection zone to the corona.}
{The code is subjected to a number of validation tests, among them the Sod shock tube
  test, the Orzag-Tang colliding shock
  test, boundary condition tests and tests of how the code treats magnetic field advection, chromospheric radiation,
  radiative transfer in an isothermal scattering atmosphere, hydrogen
  ionization and thermal conduction.}
{\bifrost{} completes the tests with good results and shows
  near linear efficiency scaling to thousands of computing cores.}
{}
\keywords{ Magnetohydrodynamics (MHD) - Radiative transfer - Methods:
  numerical - Sun: atmosphere - Stars: atmospheres}
\begin{document}

\maketitle

\section{Introduction}
The development of faster computers and better algorithms has made simulations a
viable experimental tool to understand a number of astrophysical
problems in detail. This development was clear more than a decade ago
\citep{1999ASSL..240.....M}.  The years that have followed have borne
this out fully and there are now a number of groups that are
modeling the outer layers of cool stars, including magnetic fields, to
a high degree of realism 
\citep{Gudiksen+Nordlund05a, 
          2004IAUS..223..385H,   % Hansteen St Petersburg IAUS 2004
          2006ApJ...642.1246S,    % Stein & Nordlund 2006
          Hansteen+etal07,          % Coimbra
          2005ESASP.596E..65S,    % Schaffenberger et al 2005 CO5BOLD
          2005A&A...429..335V,    % Vogler MuRAM 2005
          2007ApJ...669.1390H,    % Heinemann et al.
          2007ApJ...665.1469A,     % Abbett 
          2008ApJ...679L..57I}.       % ...and Isobe

Vital to this effort was the development of multi-group techniques to handle
radiative transfer in the photosphere \citep{Nordlund:1982} and for
models extending into the chromosphere, scattering \citep{Skartlien:2000}. The majority of the codes mentioned above
employ these multi-group methods. 

As the initial exploratory phase for codes including
magnetoconvection is nearing successful
completion, a number of challenging problems are now being
considered with some confidence. 

The problems include issues such as
the existence and formation of supergranulation,
\citep{2009ASPC..416..421S,2010AAS...21621103S}    % Stein etal Deep run
the appearance of faculae,
\citep{2004ApJ...607L..59K,     % Keller et al
          2004ApJ...610L.137C}    % Carlsson et al
the formation of active regions and spots
\citep{Cheung+etal07,             % Active region
          2007ApJ...669.1390H,    % Heinemann et al.
          2009Sci...325..171R}      % Rempel Sunspot 2009
the flux emergence into the chromosphere and corona
\citep{2008ApJ...679..871M,Martinez-Sykora:2009rw,    % Juan Flux emergence
          2009A&A...507..949T}    % Fernando Flux emergence 
the structure and heating of the chromosphere and corona
\citep{2007ApJ...665.1469A,     % Abbett
          2010MmSAI..81..582C,   % Mats fra Sac Peak
          2010ApJ...718.1070H}    % Viggo om Red-shifts....
% this example could perhaps be dropped? If not, maybe we could mention the asymmetry paper too Martinez-Sykora et.al 2011?
and acceleration of spicules
\citep{2006ApJ...647L..73H,2007ApJ...655..624D,      % Dynamic fibrils in 2d
          Heggland+etal07,      % Heggland et al.
          2009ApJ...701.1569M,    % Juan, 3d spicules
          2010arXiv1011.4703M}.   % Juan, Spicule type II
%         Brun+Palacios09 not really outer layer of cool stars....
%         Ludwig+etal09    why this one? 

Recent codes solve the full radiative magnetohydrodynamic (MHD) 
equations with some precision. However,
additional physics may have to be considered in order to solve problems
inherent to the low density, low and/or high temperature conditions of
the outer solar atmosphere. These effects encompass the inclusion of
thermal conduction by various methods 
\citep{Gudiksen+Nordlund05a,Hansteen+etal07}, but also
Generalized Ohm's law \citep{2007ApJ...666..541A}, as well as 
non-equilibrium effects such as hydrogen ionization and
molecule formation 
\citep{Leenaarts+etal07,          % 
          2007A&A...462L..31W},   % Sven and CO
are now being added to these codes.

Numerical simulations require complex algorithms to solve
the physics required, but in addition, combining high spatial
resolution with the large spatial scales characteristic of atmospheric
phenomena requires large memory and many CPU hours computing time. 
The cost of high performance computing (HPC)
specific hardware has driven the market for  supercomputers to be 
focused mainly on utilizing relatively cheap off-the-shelf computer
parts instead of developing specific supercomputing hardware. The
cheapest performance can be gained from connecting almost standard
computers through a local network. The interconnect speed and
communication software implementation is what sets one cluster apart
from another. Such clusters have a distributed memory architecture,
which is different from previous generations of supercomputers with
specifically built hardware which were typically shared memory vector
machines. The consequence is that the numerical codes used in for
instance astrophysics have to be highly parallelized in order to
perform well on large HPC systems using, for example, the 
Message Passing Interface (MPI). In the near future we expect further 
developments of these and other techniques such as the widespread use
of adaptive mesh refinement and graphics processor units (GPUs).

In this paper we present a code, \bifrost{}\footnote[1]{In Norse mythology, {\bifrost} (pron. bee-frost) is the name of the rainbow bridge from Midgard (the real of man) to Asgard (the realm of the gods), build from fire, water and air.}
to solve the MHD equations in a stellar atmosphere context,
specifically designed to
take advantage of the environment provided by modern massively
parallel machines connected through potentially slow communication
channels. The main design goal has been to reduce the amount of
communication required at each timestep while retaining good scaling
on problems requiring up to several thousand processors. 

In addition, we aim to create a flexible design in which various physical processes and extensions to the basic MHD equations are straightforward to implement without rewriting large portions of the code.

\section{Basic equations and set-up}
\bifrost{} is built on several generations of previous numerical codes,
where the Oslo Stagger Code is the latest, which at their core have the
same implemented method \citep{Nordlund+Galsgaard95,galsgaard+nordlund96}. The core of the
code remains the same, but as the previous generations of codes
required shared memory architectures, the need for a new massively
parallel code able to run on distributed memory computers was
obvious. As the core of the code was rewritten, we had the opportunity
to make  \bifrost{} much more modular and user friendly than the previous
generations of codes.

At the core of the code is a staggered mesh explicit code that solves
the standard MHD partial differential equations (PDEs) on a Cartesian grid:  
\bean
\ddt{\rho}&=&-\Div \rho \vec{u}\label{eq:mass-cons}\\
\ddt{\rho \vec{u}}&=&-\Div \left(\rho \vec{u}
\vec{u}-\tau\right)-\Grad P+\vec{J}\times\vec{B}+\rho \vec{g}\\
\mu \vec{J}&=&\Rot{\vec{B}}\\
\vec{E}&=&\eta\vec{J}-\vec{u}\times\vec{B}\\
\ddt{\vec{B}}&=&-\Rot{\vec{E}}\label{eq:induction}\\
\ddt{e}&=&-\Div e\vec{u}-P\Div \vec{u} + Q\label{eq:eqofenergy}
\eean
\noindent
where $\rho, \vec{u}, e, \vec{B}$  are the density, the
velocity vector, the internal energy per unit volume and the magnetic
flux density vector respectively. $\matrix{\tau}, P, \vec{J}, \vec{g},
\mu, \vec{E}$ and  $\eta$ are the stress tensor, the gas pressure, the
electric current density vector, the gravitational acceleration, the
vacuum permeability, the electric field vector and the magnetic
diffusivity respectively. The quantity $Q$ can contain a number of
terms, depending on the individual experiment. It could for instance
contain a term from the chosen Equation Of State(EOS), a term containing
the effect of the Spitzer thermal conductivity, a term from radiative
transfer, etc. The EOS needed to close this set of
equation can be anything from a simple ideal gas EOS to
a complex EOS including detailed microphysics (See \Sec{sec:eos}). 

\section{The method} \label{sec:method}
The basic operator in the code is the 6th order differential operator. The derivative of the function $f$ in the grid point location $+\nicefrac{1}{2}$ has the form
\bean
\ddx{f_{+\nicefrac{1}{2}}} &=& a_{\vec{x}} \left[ f_{+1}-f_{\, 0\;\, }\right]+ \nonumber \\
&&b_{\vec{x}}\left[f_{+2}-f_{-1}\right]+\nonumber\\
&&c_{\vec{x}}\left[f_{+3\,}-f_{-2}\right]
\eean
where the subscript is the gridpoint number and $a_x,b_x,c_x$ depend on the grid point distance. 
As the derivative operator produces results that lie half way between
grid points, the variables are not co-located in space, but placed on
staggered grids such that some variables are placed at cell centers,
some on cell faces and some at cell corners. By carefully choosing
which variables to place in each of these locations, it is possible to
minimize the number of interpolations needed in order to realign
computed values in space. Nevertheless, it is not possible to escape interpolations, and when it is needed the interpolation procedures used are 5th order and look very similar to the derivative operators: 
\bean
f_{+\nicefrac{1}{2}}&=&a\left[f_{\, 0\;\,}+f_{+1}\right]+\nonumber\\
&&b\left[f_{-1}+f_{+2}\right]+\nonumber\\
&&c\left[f_{-2\,}+f_{+3}\right]
\eean
where $a,b$ and $c$ are constants.

The computational grid can be stretched in one direction at a time, meaning that $d\vec{x}$ is not constant in the computational volume. The stretching of the grid is done by a simple Jacobian transformation after a derivative operator is used. Stretching in more than one direction cannot be accomplished in this simple manner without using a number of interpolations which would decrease precision, and at the moment it is not possible to employ an adaptive mesh refinement scheme in the code, but due to the modularity of the code this can be implemented at a later stage. 

\subsection{Diffusion}
All numerical codes are diffusive in nature, just due to the discrete
nature of the algorithms used in solving the equations and because of
the inaccurate machines used to solve them on. This is even true for
implicit codes that do
not contain any direct diffusive terms in their equations, that said,
the diffusion in implicit codes is significantly smaller than for
explicitly diffusive codes of 
the same order. \bifrost{} is an explicit code and it is therefore
necessary to include diffusive terms in the \Eqs{eq:mass-cons}{eq:eqofenergy} in order to maintain
stability. \bifrost{} employs a diffusive operator which is split in two major parts: A small global diffusive term and a location-specific diffusion term (sometimes dubbed ``hyper diffusion''). The diffusive operator in one dimension is of the form
\begin{equation}
\frac{\partial f}{\partial t} = \ldots +\frac{\partial}{\partial x}\left[\nu \; dx\; \left(\nu_1 C_\mathrm{fast} + \nu_2 |\vec{u}| + \nu_3\; dx\; \nabla_{x}^1 u_x \right) \frac{\partial f}{\partial x} Q\,\left(\frac{\partial f}{\partial x }\right) \right]\\
\end{equation}
where
\begin{equation}
Q\left(g\right) = \frac{\ddx{}\left(\left|\ddx{g}\right|\right)}{|g|+\frac{1}{q}\ddx{}\left(\left|\ddx{g}\right|\right)}
\end{equation}
and $\nabla^1_x$ is the first order gradient in the $x$ direction. 
The splitting of the diffusive terms into local and global components
makes it possible to run the code with a global diffusivity that is at
least a factor 10 less than if the global term were the only one
implemented in the code. The splitting also makes it impossible to
provide a single Reynolds number, Magnetic Reynolds number or any
other dimensionless number that includes the diffusion constant, since
diffusion is not constant in space or time when running
\bifrost{}. Therefore it is only possible to provide a range of the above
mentioned numbers, of which the smaller, global, value for the
diffusion would be correct for most of the simulation volume most of
the time (unless, for instance, simulating supersonic turbulence),
while the higher number of the range for the diffusion would be valid
in locations characterized by very large values of the gradients of a 
certain variable. Thus, in principle, the code can be run at much
greater values of the relevant Reynolds numbers in most of the
computational domain than would be otherwise possible using a single
diffusion coefficient.
 
 \section{Modules}
\bifrost{} has been created with a high degree of modularity. The code
has a basic skeleton which connects to a number of modules. Any of
these modules can contain a simple procedure or method, include a
number of sub-modules or be left empty.  For example, the timestepping
procedure can be swapped between a Runge-Kutta scheme and a Hyman
scheme by changing only a single line in an input file to the
compiler, plus a recompilation. Most interestingly, the code can include any number of modules that provide new physics, or boundary conditions, in the same simple way.

As there can be several implemented modules handling the same job in
the code (timestepping, EOS, radiative transfer etc)
this section contains the most important of these, which are presented
either with a short description if the modules are standard numerical
schemes, or in more detail if they are specific to \bifrost{}. 

\section{Timestepping}
Timestepping can be handled by two different procedures: A third-order
Runge-Kutta method or a third-order Hyman method. Both of these
procedures produce nearly identical results. The Runge-Kutta method is
able to handle a longer timestep, while being more computationally
intensive, so in terms of CPU time their effectiveness is almost the same. 
\subsection{Third-order Runge-Kutta timestepping scheme}
The implemented 3rd order Runge-Kutta scheme splits the timestep into
three sub-steps. In order to take one timestep all the partial
differential equations are solved three times but it is not necessary
to save more than two results at a time in the same timestep, making
this scheme $2N$ in memory requirements (where $N$ in the amount of
memory needed to run through the partial differential equations
once). For large simulations that can be a considerable memory saving
feature. The advantage in this method is that the intermediate
timestep results can be used to extrapolate the result of the total
timestep further in time than that possible by using three separate
time steps, while at the same time having a high order precision. 
The Runge-Kutta method is defined by assuming the change in the
variable $f$ can be written 

\begin{equation}
\frac{\partial f}{\partial t} = F\left(t,f(t)\right)\textrm{\ .}
\end{equation}
The change during one timestep $\Delta t$ of the variable $f$ is then given by 
\begin{equation}
f(t+\Delta t)=f(t)+\frac{\Delta t}{6}\left(k_1+4k_2+k_3\right)\textrm{\ ,}
\end{equation}
where
\begin{eqnarray}
k_1&=&F\left(t,f(t)\right) \textrm{\ ,} \\
k_2&=&F\left(t+\frac{1}{2}\Delta t ,f(t)+\frac{1}{2}k_1\Delta t \right) \textrm{\hspace{0.5cm} and} \\
k_3&=&F\left(t+\Delta t,f(t)-k_1\Delta t +2k_2\Delta t\right) \textrm{\ .}
\end{eqnarray}

\subsection{Third-order Hyman timestepping scheme}
The third-order Hyman predictor-corrector scheme is described by
\cite{Hyman79}. It is an iterative multistep method
employing a leap-frog scheme to attain 3rd order accuracy in time. The method is quite simple and can be described by assuming that the differential equations can be written in the following form: 

\begin{equation}
\frac{\partial f}{\partial t} = F\left(t,f(t)\right) \label{eq:simple_hyman}
\end{equation}
The Hyman method's first step is to find a second order predictive solution to \Eq{eq:simple_hyman} by using the formula: 
\begin{equation}
f_{n+1}^{(2)} = (1-r^2) f_n+r^2f_{n-1}+\Delta t (1+r) F_n
\end{equation}
where the superscript is the order of the term, the subscript is the timestep number and $F\left(t,f_n(t)\right)$ has been written $F_n$ for simplicity and 
\begin{equation}
r\equiv(\Delta t_{n+1}/\Delta t_{n})
\end{equation}
with $\Delta t$ being the timestep. Then the PDE's are solved again and finally a corrector is applied given by 
\begin{eqnarray}
F_{n+1}^{(3)}&=& \left[(2-r)(1+r)^2 f_n+r^3f_{n-1} + \right. \nonumber \\
&&\left. \Delta t (1+r)^2F_n+ \Delta t(1+r)F_{n+1}^{(2)}\right]/(2+3r)\textrm{\ .}
\end{eqnarray}

\section{Boundary conditions}
\bifrost{} is able to make the computational box periodic in one, two or
three dimensions. Non-periodic boundaries imply that a boundary condition must be imposed on that boundary. \bifrost{} implements non-periodic boundary conditions by padding the computational domain with ghost zones in the relevant direction and filling them according to the boundary condition chosen. Several standard boundary conditions are implemented including symmetric, antisymmetric 
as well as extrapolated boundary values that fill the ghost zones
before the MHD PDEs are computed. Experiment-specific boundary
conditions such as constant entropy flux for stellar convection
simulations and characteristic boundary conditions can also be used by
adhering to a simple format for boundary calls. In the latter case of
characteristic boundaries the
conservative MHD PDEs are replaced by the characteristic PDEs at the boundary
zones before the variables in these zones are updated in the usual manner.

\subsection{Characteristic boundary conditions}
The characteristic boundary conditions aim at transferring
disturbances through the boundaries without any or at least with minimal
reflection, a problem that can plague standard boundary conditions
that simply use symmetric or extrapolated values.
In this case the 8 MHD equations are written in terms of the
characteristics and split into horizontal and vertical
components in the following manner. 

With $\UU^{'}$ defining a vector that contains the conserved
MHD variables we may write the equations as 
\begin{equation}
\ddt{\UU^{'}}+\sum_{k=1}^m \ddxk{\FF^k}=\DD^{'}
\textrm{\ ,}
\end{equation} 
where $\FF$ contains the fluxes, $\DD^{'}$ the source terms, and $m=3$ for a 3D problem.
These conservation equations can be transformed using linear algebra into ``primitive'',
wave-like equations for a corresponding set of 8 primitive
variables $\UU$,
\begin{equation}
\ddt{\UU}+\sum_{k=1}^m {\AM^k} \ddxk{\UU}=\DD
\textrm{\ ,}
\end{equation}
where the three $\AM^k$ are $8 \times 8$ matrices. 
The choice of primitive variables $\UU$ is not unique, and in
principle could be taken to be the conserved variables. The goal of this
procedure is then to arrive at equations that resemble simple advection
equations where specifying boundary conditions is straightforward:
extrapolation based on one-sided derivatives for outgoing
characteristics and on the basis of exterior data, such as no incident
waves, for incoming characteristics.
By combining all the flux divergence terms except those in the direction perpendicular to the boundary, now called $z$, with the source term (forming a new source term $\CC$) we can write
characteristic equations for the
$z$ direction in the sought after form
\begin{equation}
\label{eq:characteristic_eqn}
\SM^{-1}\ddt{\UU}+{\mathbf{\Lambda}}\SM^{-1}\ddz{\UU}=\SM^{-1}\CC
\textrm{\ ,}
\end{equation}
as shown by e.g. \citet{Thompson19871}. Here, the rows $i$ of the matrix
$\SM^{-1}$ are given by the left eigenvectors ${\mathbf{l}}_i^T$, and
$\mathbf{\Lambda}$ is the diagonal matrix formed by the eigenvalues
$\lambda_i$ of the matrix $\AM^k$ belonging to the $z$
direction. We now define a vector $\mathbf{d}$ containing the $z$
derivatives of the characteristic equations as 
\[
{\mathbf{d}}\equiv\mathbf{\Lambda}\SM^{-1}\ddz{\UU}
\textrm{\ .}
\]
Having isolated the various characteristic wave modes propagating in
the $z$ direction we left-multiply
\Eq{eq:characteristic_eqn} by $\SM$ in order to write the
MHD equations in primitive form in terms of $\mathbf{d}$:
\begin{equation}
\label{eq:primitive_char_eqn}
\ddt{\UU} + {\SM}\mathbf{d}  =  \CC
\end{equation}
where the primitive variables $\UU$ are comprised of the variables $\rho$, $\uu$, $e$ and
$\BB$.  The $\mathbf{d}$ vector containing the $z$ derivatives $d_1$ -- $d_8$ 
constitute the information that is flowing along the characteristics;
outflowing characteristics are defined by the interior solution,
inflowing characteristics by the requirement that the incoming wave be
constant in time. Expressions for \Eq{eq:primitive_char_eqn} in primitive form and for the characteristic derivatives $\mathbf{d}$ can be found in Appendix \ref{app:char_bound}. 

\section{Equation of state}\label{sec:eos}

The code provides several different EOS modules,
which can be chosen according to the experiment one wants to
perform. The EOS modules provide the temperature and pressure and
their thermodynamic derivatives as a function of mass density and
internal energy per mass. There are currently three different modules.

The first implements an ideal gas EOS, suited for testing and
idealized experiments.

The second module implements an EOS based on tables generated with the
Uppsala Opacity Package \citep{gustafsson+etal75}. It assumes local thermodynamic equilibrium (LTE) 
for atomic level populations and instantaneous molecular dissociation
equilibria. This package is required when running with full radiative
transfer (see \Sec{Sec:RadiativeTransfer_full}), as it also provides the opacity, thermal emission and
scattering probability for the radiation bins. The tables are
generated with a separate program; different tables can be generated
to account for different stellar spectral type, chemical composition and number of radiation
bins.

The third package computes the gas temperature, gas pressure and
electron density explicitly based on the non-equilibrium ionization of
hydrogen in the solar atmosphere. This package can only be used for
simulations of the solar atmosphere, as it depends on a number of
parameters that vary with stellar spectral type. These parameters have
so far only been determined for the sun. More details on
non-equilibrium hydrogen ionization are given in \Sec{sec:hion}.

\section{Radiative transfer}\label{Sec:RadiativeTransfer}
Full 3D radiative transfer is computationally very costly. The
properties of radiation are very different from the strictly local
problem of MHD, since radiation can couple thermodynamic properties
over arbitrarily long distances. For a code that relies on the local
properties of MHD to parallelize, this makes radiative transfer costly
not just in shear computations needed to solve the problem, but also
because the results of the computation must be communicated to all
nodes. As a result, the modules handling radiative transfer in \bifrost{}
have employed assumptions which simplify the calculations to some
extent. At the moment three modules handle radiative
transfer depending on the problem at hand and they are often combined. 

The electron number density is often needed in the radiative transfer
modules (e.g. for the calculation of collisional excitation rates and 
opacities). The electron number density computation 
depends on which EOS package is used: if hydrogen non-equilibrium
ionization is calculated, the electron number density comes from the
simultaneous solution of the hydrogen rate equations, the energy
equation and the charge conservation equation (see \Sec{sec:hion} for
details). For the EOS package based on tables generated with the Uppsala
Opacity Package, the electron density comes from solving the Saha
equation for all species. If the electron density is not provided by the
EOS package but is needed (\eg\ for the chromospheric radiation approximation,
see \Sec{sec:genrad}), it is
computed using the Saha relation for hydrogen, but setting a floor to the
ionization degree of $10^{-4}$ to account for the easily ionized
metals in the solar atmosphere. 

\subsection{Optically thin radiative transfer}\label{sec:thinrad}
In the outer solar atmosphere (and many outer stellar atmospheres)
radiative losses can be simply treated assuming that the atmosphere is
optically thin. In the sun this is true for most lines from the upper
chromosphere/lower transition region up to the corona. In this case
(and assuming ionization equilibrium only dependent on temperature)
the radiative transfer problem reduces to a radiative loss which
only depends on density and temperature of the form (note that positive Q corresponds to heating):
\bean
Q_{\rm thin}= - n_{\rm H} \; n_{\rm e} f(T) \exp({-P/P_0})
\eean
where $n_{\rm H}$ and $n_{\rm e}$ are the number densities of hydrogen
and electrons respectively,  $f(T)$ is a function of the
temperature and the term $\exp({-P/P_0})$ provides a cutoff where
$P>P_0$. The radiation stems mainly from the resonance lines of
multiply ionized elements such as carbon, oxygen, and iron and the
function $f(T)$ can be pre-computed assuming
ionization equilibrium. This relation is valid from roughly $2\times
10^4$ K and up. In \bifrost{} the function $f(T)$ is computed by using
ionization and recombination rates given by
\citet{Arnaud+Rothenflug85} and \citet{shull+vansteenberg82} and
collisional excitation rates given by the HAO-DIAPER atom data package 
\citep{judge+meisner94} including lines from He, C, O, Ne and Fe. The
optically thin losses from hydrogen lines are normally calculated in the
chromospheric approximate radiation part, see \Sec{sec:genrad}, but
may instead be included here in the cases where a calibration of the
chromospheric radiative losses is lacking.
The total hydrogen number density is derived from the plasma density $\rho$
assuming solar abundances. We set $P_0$ to a typical value of the
mid-chromospheric pressure and the term $\exp({-P/P_0})$ then makes
sure that there are no optically thin contributions calculated in the deep
convection zone where the temperature is also above $2\times 10^4$ K.

\subsection{Chromospheric radiation approximation} \label{sec:genrad}
Chromospheric radiative losses are dominated by a small number of
strong lines from hydrogen, calcium and magnesium \citep{VALIII}.
The source functions
and opacities of these lines are very much out of local equilibrium and
the optical depth is significant; neither the optically thin approach
of \Sec{sec:thinrad} nor the full radiative transfer with coherent
scattering and LTE opacities (see \Sec{Sec:RadiativeTransfer_full})
give good results. We approximate the radiative loss in
these lines with the formulae
\bean
Q_\mathrm{[H,Ca,Mg]} = - C(T)_\mathrm{[H,Ca, Mg]} \nel \rho
\phi_\mathrm{[H,Ca, Mg]} (m_{\rm c})
\eean
where $C(T)\nel\rho$ gives the total collisional
excitation rate (in energy per volume per unit time),
$\phi(m_{\rm  c})$ gives the probability that this energy escapes 
from the atmosphere and $m_{\rm c}$ is the column mass. 
The function $\phi(m_{\rm  c})$ and the temperature-dependent coefficient $C(T)$ are
determined for each element from detailed 1D radiative transfer computations with the
RADYN code
\citep[see, \eg][]{1995ApJ...440L..29C} % Carlsson+Stein Radyn
and 2D computations with Multi3D
\citep{2009ASPC..415...87L}. % Leenaarts&Carlsson Multi3d
These functions include hydrogen lines and
the Lyman continuum and all lines and continua from \CaII\ and \MgII.
The method is described in detail in Carlsson \& Leenaarts (in preparation).
%\citet{Carlsson+Leenaarts2011}.

Half of the UV radiation lost from the corona in optically thin lines (see
\Sec{sec:thinrad}) goes towards the sun and most of that is 
eventually absorbed in the chromosphere, mostly in the Lyman continuum
and the \ion{He}{I} continuum. This radiative heating is modeled through the representative bound-free absorption
cross-section $\sigma$ of \ion{He}{I} at 25~nm:
\bean \label{eq:QHe}
Q_\mathrm{He} = \sigma_\mathrm{He,25\,nm} \, n_\mathrm{He~I} \,
\exp({-\tau_{\mathrm{He,25\,nm}}}) \, J_\mathrm{thin}
\eean
with $J_\mathrm{thin}$ the angle-averaged radiation field caused by
$Q_\mathrm{thin}$
(see Carlsson \& Leenaarts in preparation).
%\citep[See][]{Carlsson+Leenaarts2011}. % Editor does not want inprep in bibliog.
%

\subsection{Full radiative transfer}\label{Sec:RadiativeTransfer_full}

Full radiative transfer computations are required when a simulation includes the convection zone beneath the photosphere, covering optically thick regions, optically thin regions, and the transition between the two regimes. A simplified treatment using, e.g., Newtonian cooling or the diffusion approximation, cannot provide sufficiently realistic radiative heating and cooling rates in this boundary layer.

Owing to the very short time scales of photon interaction and propagation in a convective stellar atmosphere, it is possible to solve radiative transfer as a time-independent problem, resulting in the expression
\bean
\mathbf{\hat{n}\cdot\nabla}I_{\lambda}(\mathbf{x},\mathbf{\hat{n}})=-\chi_{\lambda}(\mathbf{x})I_{\lambda}(\mathbf{x},\mathbf{\hat{n}})+j_{\lambda}(\mathbf{x},\mathbf{\hat{n}})
\textrm{\ ,}
\label{eqn:radtrans}
\eean
where $I_{\lambda}(\mathbf{x},\mathbf{\hat{n}})$ is the monochromatic specific intensity at spatial point $\mathbf{x}$ for a beam in direction $\mathbf{\hat{n}}$ at wavelength $\lambda$, $\chi_{\lambda}$ is the gas opacity, and $j_{\lambda}$ is the emissivity. The two material constants $\chi_{\lambda}$ and $j_{\lambda}$ depend both on the thermodynamic state of the gas and on the radiation field, and are highly wavelength-dependent in the presence of spectral lines and other atomic and molecular transitions. Velocity fields, such as convective motions in the atmosphere, lead to an additional coupling between ray directions and wavelengths through Doppler shifts. The complexity of taking full account of the underlying physical mechanisms is computationally prohibitive. We therefore assume a static medium and LTE for the gas opacity, which thus depends only on the local gas temperature and the gas density, and which can therefore be precomputed and tabulated. \citet{Skartlien:2000} and \citet{Hayeketal:2010} showed the importance of photon scattering in simulations of the higher solar atmosphere; we therefore include a coherent scattering contribution in the gas emission, which requires an iterative solution of the radiative transfer equation to obtain a consistent radiation field.

The vast number of spectral lines encountered in a stellar atmosphere requires, in principle, the solution of a large number of radiative transfer problems. This is currently too demanding for realistic 3D radiation-hydrodynamical calculations. We therefore approximate the opacity spectrum by substituting the monochromatic $\chi_{\lambda}$ with a small number of mean opacities \citep{Nordlund:1982,Skartlien:2000} and solving wavelength-integrated radiative transfer.

The radiation field encountered in cool stellar atmospheres does not contribute significantly to the momentum balance. We consider only a radiative heating rate $Q_{\mathrm{rad,i}}$ for every mean opacity (index $i$), given by the first moment of \Eq{eqn:radtrans}:
\bean
Q_{\mathrm{rad},i}=-\mathbf{\nabla\cdot F_{i}}=4\pi\chi_{i}(J_{i}-S_{i})
\textrm{\ ,}
\eean
where $\mathbf{F_{i}}$ is the radiative energy flux, $J_{i}$ is the mean intensity and $S_{i}\equiv j_{i}/\chi_{i}$ is the source function. The solver computes the radiation field using the short characteristics technique in every spatial subdomain and iterates the solution until convergence. The radiative heating rates are then added to the energy equation (\Eq{eq:eqofenergy}). If the hydrodynamical timesteps are sufficiently short that changes in the radiative energy flux are small, full radiative transfer only needs to be computed in a fraction of the hydrodynamical timesteps. A detailed description of the numerical methods and the parallelization techniques used for the full radiative transfer module in \bifrost{} can be found in 
\citet{Hayeketal:2010}.

\section{Hydrogen ionization} \label{sec:hion} 
The ionization of hydrogen in the solar chromosphere does not obey LTE
or statistical equilibrium
\citep{Carlsson+Stein02}. % Carlsson & Stein 1D RADYN
Proper modeling of this ionization requires that non-equilibrium
effects be taken into account. 

The hydrogen ionization module solves the time-dependent rate
equations for the atomic hydrogen level populations $n_i$
\begin{equation} \label{eq:hionevol}
 \frac{\partial n_i}{\partial t} + \nabla \cdot (n_i\vec{u}) =
 \sum_{j,j \ne i}^{n_{\rml}} n_j P_{ji} - n_i \sum_{j,j \ne i}^{n_{\rml}} P_{ij}
\end{equation}
together with an equation for time-dependent H$_2$ molecule formation
and equations for energy and charge conservation. Here $P_{ij}$ is the
rate coefficient for the transition from level $i$ to $j$ and $n_{\rm l}$ is the number of energy levels in the hydrogen atom (normally set to 6 in \bifrost{}). Solution of
this system of equations yields the gas temperature, electron density
and the atomic and molecular hydrogen populations (\nhtwo). The radiative transition
rate coefficients in the rate equations are prescribed as determined by 
\citet{sollum99}.
This removes the effect of the global coupling of the radiation and
makes the problem computationally tractable, but still computationally
demanding. 

The gas pressure is computed as
\begin{equation} \label{eq:hionpressure}
P = k T \left( \nel + \nhtwo + \sum_{i=1}^{n_{\rm l}} n_i + \no \right)
\textrm{\ ,}
\end{equation}
with $\no$  the number density of all other atoms and molecules that are not,
or do not contain, hydrogen. 

Full  non-LTE radiation tables
as functions of $\nel$ and $T$ can be used to replace the radiation tables
as functions of $e$ and $\rho$, which assume hydrogen ionization based
on Saha ionization equilibrium.

The method is described in detail in
\citet{Leenaarts+etal07},
the additional equation for time-dependent \htwo$\,\!$  formation is given
in Appendix~\ref{app:hion}.

\section{Thermal conduction}

As the plasma temperature rises towards one million degrees in the tenuous 
transition region and corona, thermal conduction becomes one of the major
terms in the energy equation, and modeling of this term is vital if this 
portion of the atmosphere is to be simulated with any fidelity. The implemented form of the thermal conduction takes the form \citep{Spitzer56} :
\begin{equation}
{\bf{F}}_\mathrm{c}=-\kappa_0T^{5/2}\nabla_{\parallel}{T}
\end{equation}
where the gradient of $T$ is taken only along the magnetic
field  ($\nabla_{\parallel}$) and $\kappa_0$ is the thermal conduction coefficient. 
The conduction across the
field is significantly smaller under the conditions present in the
solar atmosphere and is smaller than the numerical diffusion, so it is
ignored. Since thermal conduction is described by a second order diffusion operator,
this introduces several difficulties: The Courant condition for a diffusive 
operator such as that scales with the grid size $\Delta \vec{x}^2$ instead of with
 $\Delta \vec{x}$ for the magnetohydrodynamic operators. This severely limits the 
time step $\Delta t$ the code can be stably run at. We have implemented two 
solutions to this problem. 

In the first, the thermal flux is calculated and if the divergence of
the thermal flux sets severe restraints on the timestep, it is
throttled back by locally lowering the thermal conduction coefficient $\kappa_0$. This method
is only acceptable when the number of points where the conduction has to be throttled
back is very small and the results must be analyzed carefully. 

A second solution is to proceed by operator splitting, such that the operator advancing 
the variables in time is $L=L_\mathrm{hydro}+L_\mathrm{conduction}$. The conductive part 
of the energy equation is handled by discretizing

\begin{equation}
{\partial e\over\partial t}=-\nabla\cdot{\bf F}_\mathrm{c}
=-\nabla_\parallel\cdot\left[\kappa_0 T^{5/2} \nabla_\parallel T\right]
\label{eq:thermal-operator}
\end{equation} 
and solving the resulting problem implicitly.

We discretize the $L_{conduction}$ operator using the Crank-Nicholson
(or `$\theta$') method 
and thus write
\begin{equation}
(e^{n+1}-e^{*})/dt=\theta\nabla_\parallel\cdot{\bf F}_\mathrm{c}^{n+1}+(1-\theta)\nabla_\parallel\cdot{\bf F}_\mathrm{c}^{n})
\label{eq:thermal-operator-discrete}
\end{equation}
\noindent where the quantities with superscript $n$ are computed before the 
hydrodynamic timestep, the quantities with superscript $*$ are computed after 
the 
hydrodynamic timestep and the quantities with superscript $n+1$ are the 
temperatures deduced implicitly. The variable $\theta$ is set to a
value between $0.5$ and 1. The implicit part of the problem is in our case
computed  using a multi-grid solver 
(A concise introduction to multi-grid methods 
may be found in chapter 19 of \cite{Press+etal1992}).

In implementing the multi-grid solver, we use the 
same domain decomposition as in the magnetohydrodynamic part of the code.
The small scale residuals in \Eq{eq:thermal-operator-discrete} are 
smoothed using a few Gauss-Seidel sweeps at high resolution, before
the resulting temporary solution is injected onto a coarser grid on
which smoothing is again performed. The process
continues by Gauss-Seidel smoothing sweeps on steadily coarser grids until 
the size of the problem on individual processors is small enough to be 
communicated and stored on each and every processor. At this point the 
partial solutions are spread to all processors which continue the 
Gauss-Seidel smoothing, now on the global problem, on steadily coarser 
grids. Finally, when the coarsest grid size is reached, the solution is driven
to  convergence by performing a great number of Gauss-Seidel sweeps. 
Subsequently, the coarser solutions are corrected and then prolonged
on successively finer grids, 
first globally, but after a certain grid size is reached the temporary
solution is spread and the problem is again solved locally, using 
the same domain decomposition as in the magnetohydrodynamic part of
the code. After each prolongation, the solution is smoothed using a
few Gauss-Seidel sweeps. The entire cycle can be repeated as many
times as desired, until a converged solution is found.

\section{Tests and validation}

\bifrost{} has been extensively tested in order to validate the results
it provides. When possible, the tests are 
taken from the literature to validate the code and allow comparison with previous work. These tests were selected for their simplicity, their utility and
their challenge to the different terms in the algorithms. 

\subsection{The Sod shock tube test}

The Sod shock tube test \citep{Sod78} has become a standard test in
computational HD and MHD. It consists of  a one-dimensional flow
discontinuity problem which provides a good test of a compressible
code's ability to capture shocks and contact discontinuities within a
small number of zones, and to produce the correct density profile in a
rarefaction wave. The test can also be used to check if the code is
able to satisfy the Rankine-Hugoniot shock jump conditions, since this
test has an analytical solution.

The code is set to run a 1D problem and using the initial conditions for the Sod problem. 
The fluid is initially at rest on either
side of a density and pressure jump. The density and pressure jumps
are chosen so that all three types of flow discontinuity (shock,
contact, and rarefaction) develop. To the left, respectively right
side  of the interface we have:

\begin{eqnarray}
\rho_R &=& 1 \textrm{\ ,}\\
\rho_L &=& 0.125 \textrm{\ ,}\\
P_R &=& 1/\gamma \textrm{\ ,}\\
P_L &=& 0.125/\gamma \textrm{\ .}
\end{eqnarray}

The ratio of specific heats is chosen to be $\gamma=5/3$ on both sides
of the interface and with a uniform magnetic field perpendicular to the 1D axis of the domain 
($B_z=1$). The units are normalized, with the density and pressure in units of the density
and pressure on the left hand side of jump and the velocity in units of the sound 
speed. The length is in unit of the size of the domain and the time in units of the 
time required to cross the domain at the speed of sound.

We have run the simulation at different resolutions and with different
values of the numerical diffusion coefficient in an effort to find
the minimal diffusion the code can be run with without developing
numerical instabilities. From this test and the following ``field advection'' 
and Orzag-Tang tests we find that the diffusion parameters
should be greater than $\nu_{1}>0.02$, $\nu_{2}>0.2$ and $\nu_{3}>0.2$
to ensure stability. Other parameters,
such as the quench parameter, have also been varied in testing the
code, in order to find the best values for capturing the Sod shocks 
properly. 

\begin{figure}
 \resizebox{\hsize}{!}{\includegraphics{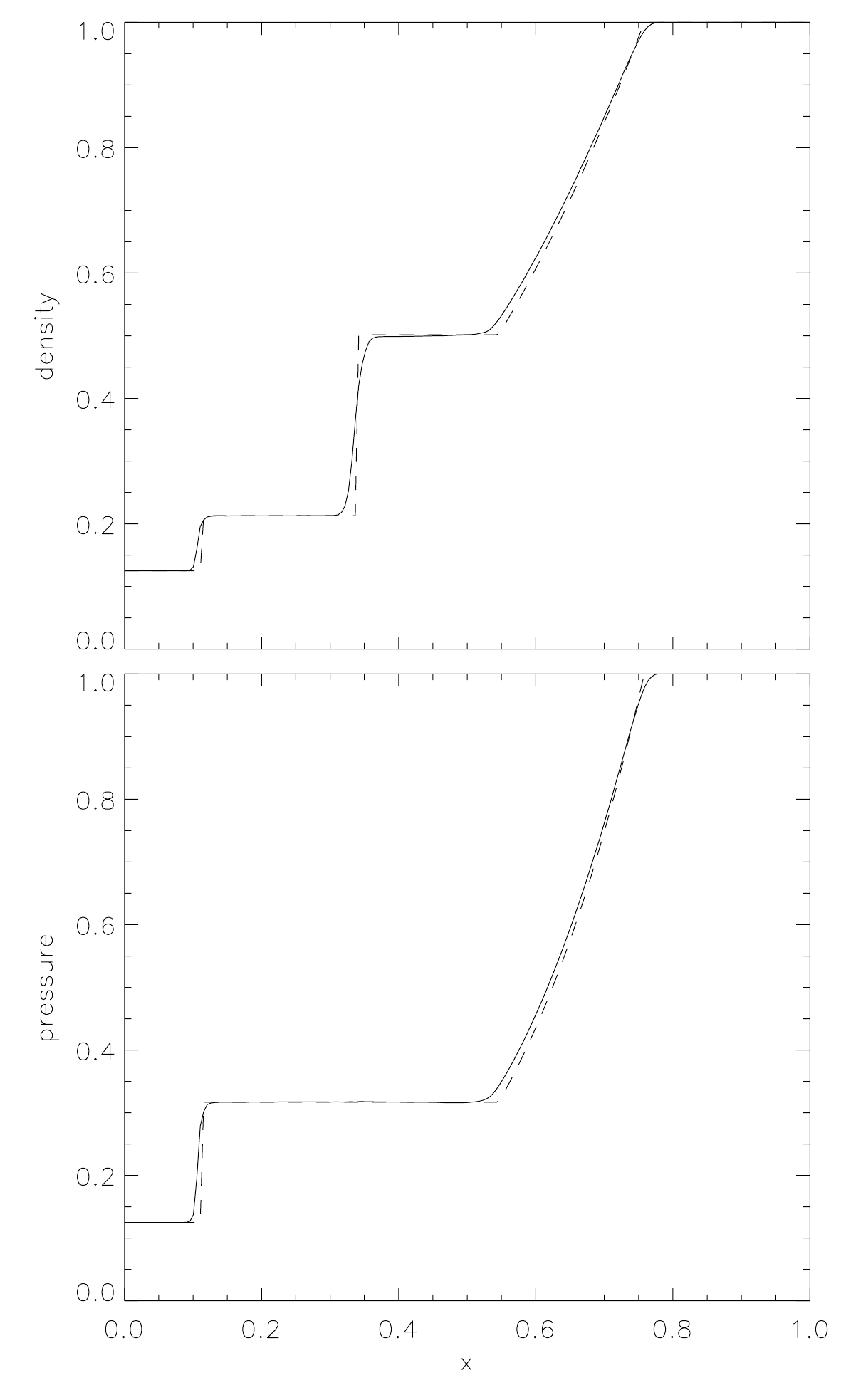}}
 \caption{\label{fig:sod}Density profile (top panel) and pressure
   divided by the maximum of the pressure profile (bottom panel) at
   time $0.193$. The solid line is the numerical result and the dashed
   line is the analytical solution.}
\end{figure}

\Fig{fig:sod} shows the density and pressure profiles at time $0.193$. 
The solid line is our 
numerical solution and the dashed line is the analytic solution at the
same instant in time. It is clear that when running with the numerical
diffusion coefficients large enough to avoid post shock numerical
instabilities, but low enough not to lose the sharp shock profiles, 
the code solves the Sod shock correctly.

\subsection{The magnetic field advection test}
This is a multidimensional convection test, which serves to test the 
conservation of the various MHD quantities such as the density, 
momentum, and magnetic field with advection. Moreover, 
multidimensional MHD problems present a special challenge to the
conservation of the divergence of the magnetic field \citep{toth+odstrcil96}. 
This can only occur if the scheme preserves $\nabla \cdot {\bf B} = 0$. 

This advection test is based on a test described previously by
\citet{DeVore91}.  The test involves advecting a cylindrical current
distribution, which forms a tube of magnetic field, diagonally across the grid. 
Any arbitrary advection angle can be chosen and the tube can have any
orientation. For the 3D results shown here, the 
problem domain is given by $0 \le x_{1} \le 1$; $-1/(2 \cos{(30^\circ)}) \le x_{2} \le 1/(2 \cos{(30^\circ)})$, 
$0 \le x_{3} \le 1$, and the flow is inclined at 60 degrees to the $x_2$ axis 
in the same plane as $x_1$; where $x_{1}$, $x_{2}$ and $x_{3}$ could be either 
$x$, $y$ or $z$. The loop is oriented in the $x_{3}$ direction. This geometry ensures 
the flow does not cross the grid along a diagonal, so the fluxes in $x_{1}$, $x_{2}$ 
and $x_{3}$-directions will be different. 

The flow velocity is set to $1.0$, with direction $u_{1}=\sin{(60^\circ)}$ 
and $u_{2}=\cos{(60^\circ)}$. We have run many tests, changing $u_x$, $u_y$ and 
$u_z$ between these $u_{1}$ and $u_{2}$ values and have checked that
there is no dependence on any direction. The density is 1, pressure is $1/\gamma$, 
and the gas constant is $\gamma = 5/3$. Periodic boundary conditions 
are used everywhere.

The magnetic field is initialized using a vector potential, to make sure that the magnetic field is initially divergence free. The potential in this example is given by:
\bea
A_{3}=0.009 e^{-\left(5r\right)^6}
\eea
where $r$ is the distance from the box centre in the $x-y$ plane, and the two other components of the vector potential are set to zero. The magnetic vector potential provides the magnetic field vector according to $\vec{B}=\Rot{\vec{A}}$, and leaves a narrow cylinder of magnetic field and the divergence of the magnetic field is zero as calculated by \bifrost{}.
 
%!The magnetic field is initialized using an arbitrary vector potential 
%!defined at zone corners; we use 
%\[
%A_{3} = \max\left(\min\left[A ( R_o - r ),2\cdot 10^{-3}\right],0\right)
%\textrm{\ .}
%\]
%The amplitude $A$ must be so small that the field is weak compared to 
%the gas pressure. A stronger field would require a more careful choice 
%of $A_{3}$ to ensure that the loop is in magnetostatic equilibrium. We use 
%$A = 10^{-3}$ with radius $R_o = 0.3$. We set up a small 
%inner radius because the stencil of the code could produce problems in 
%the center of the loop. 

%Face-centered magnetic fields are computed using \mbox{${\bf B} = \nabla \times A_{3}$}
%to guarantee $\nabla \cdot {\bf B} = 0$ initially. Note that for the vector potential 
%we have adopted, the second derivative (current density) is discontinuous. 

The code has solved this test with different setups of the direction 
of the initial velocity and orientation of the loop.  We have also
checked that the test performs successfully with varying diffusion 
and quench parameters. The increase in $\nabla \cdot {\bf B}$ with
time lies within the numerical error. This means that every $10^6$ 
time-step the cumulative error of the numerical errors makes it necessary to perform 
a cleaning of the $\nabla \cdot {\bf B}$.

\begin{figure}
\resizebox{\hsize}{!}{\includegraphics{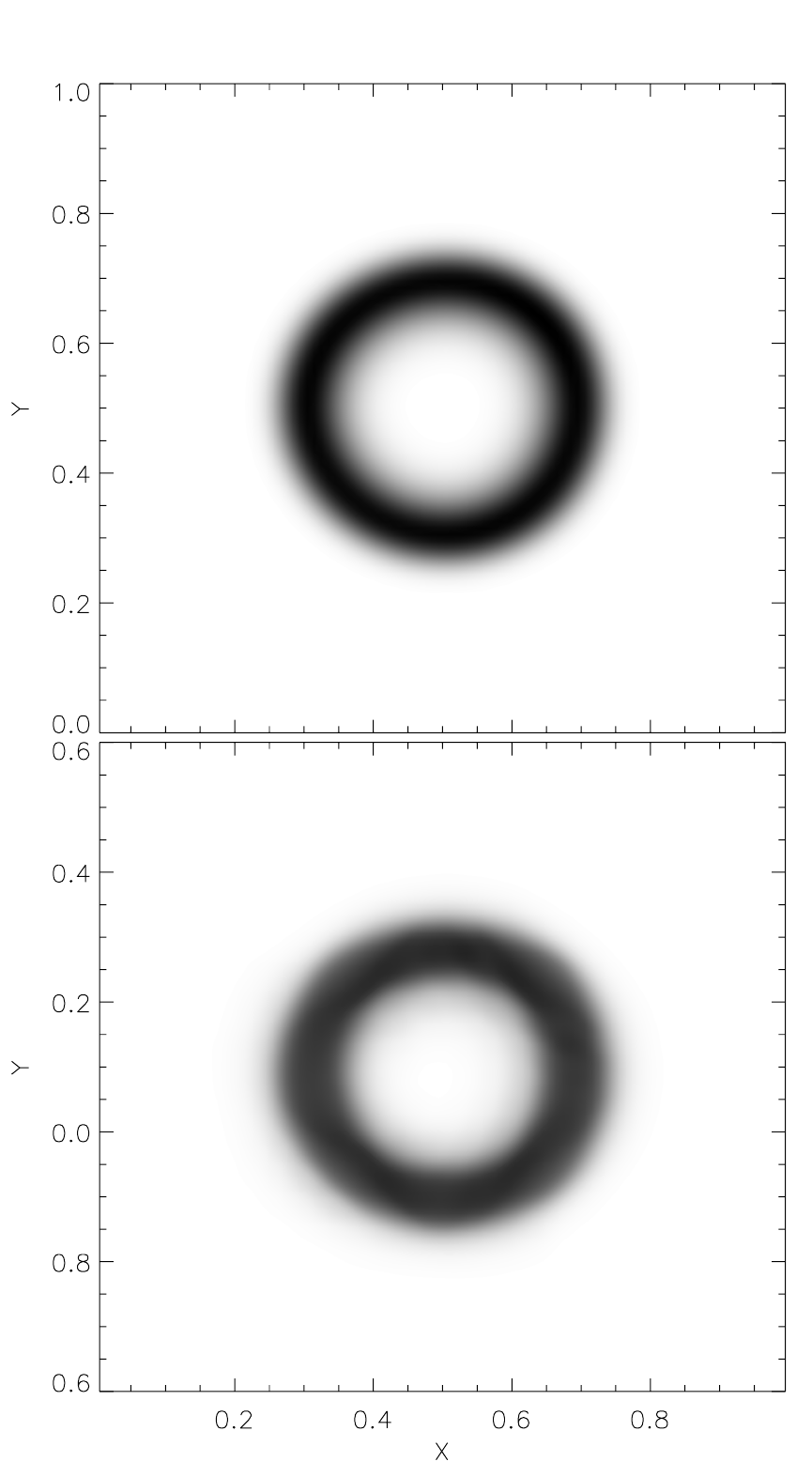}}
%  \resizebox{\hsize}{!}{\includegraphics{figs/adv1.eps}}
% \resizebox{\hsize}{!}{\includegraphics{figs/adv13.eps}}
 \caption{\label{fig:loop} The magnitude of the magnetic field shown at $t=0$ (top) and at $t=1.155$ (bottom) when the cylinder has moved one box length along the x-axis, and roughly half way along the y-axis. The lower contour plot has been centered to make a comparison easier.} 
\end{figure}

In \Fig{fig:loop}, the magnetic cylinder shows small alterations in shape due mainly to the rectangular grid used, and because the width of the magnetic ring initially was chosen to be at the limit of the resolution capability of the code, which exaggerates the effect of the grid.
%In \Fig{fig:loop} 
%we show how the pressure evolves in the case of a grid of 
%$100\times100\times100$ points. The ring has been smoothed by diffusion after several %crossings through the domain.

\subsection{The Orszag-Tang test}

The problem was first studied by \citet{Orszag+Tang79}. Since then
solutions to the problem have been extensively compared in numerical
MHD simulations. This provides a good way to compare codes. 

The Orszag-Tang (O-T) vertex is a model problem for testing the transition to
supersonic 2D MHD turbulence. Thus, the problem tests how robust the
code is at handling the formation of MHD shocks and shock-shock
interactions. The problem can also provide some quantitative estimates
of how significant magnetic monopoles affect the numerical solutions,
testing the $\nabla \cdot {\bf B} = 0$ condition. Finally, as
mentioned, the problem is a very common test of numerical MHD 
codes in two dimensions, and has been used in many previous studies. 

The set up is the following: The domain is 2D and goes from $0 \le x \le 1$ and $0 \le y \le 1$. The boundary conditions are periodic. The density $\rho = 1$, the pressure $P=1/\gamma$ and $\gamma = 5/3$ everywhere. Note that this choice gives a sound speed of $c_\mathrm{s} = \sqrt{\gamma P/\rho }= 1$. The initial velocities are periodic with:

\begin{eqnarray}
u_x &=& - \sin(2\pi y)\textrm{\ ,}\\
u_y &=& \;\;\: \sin(2\pi x)\textrm{\ .}
\end{eqnarray}
\noindent The magnetic field is initialized using a periodic vector potential defined at zone corners:
 
 \begin{eqnarray} 
 A_z &=& B_o \left[ \frac{\cos(4\pi x)}{4\pi} + \frac{\cos(2\pi y)}{2\pi}\right]
 \textrm{\ .}
 \end{eqnarray}
 
\noindent with $B_o = \sqrt{1/(4\pi)}$. Face-centered magnetic fields are computed using $B = \nabla \times A_z$ to guarantee $\nabla \cdot {\bf B} = 0$ initially. This gives:
 
 \begin{eqnarray} 
 B_x &= &- B_o \sin(2\pi y)\textrm{\ ,}\\ 
 B_y &= &\;\,\, B_o \sin(4\pi x)\textrm{\ .}
\end{eqnarray} 

\begin{figure*}
\centering
 \includegraphics[width=17cm]{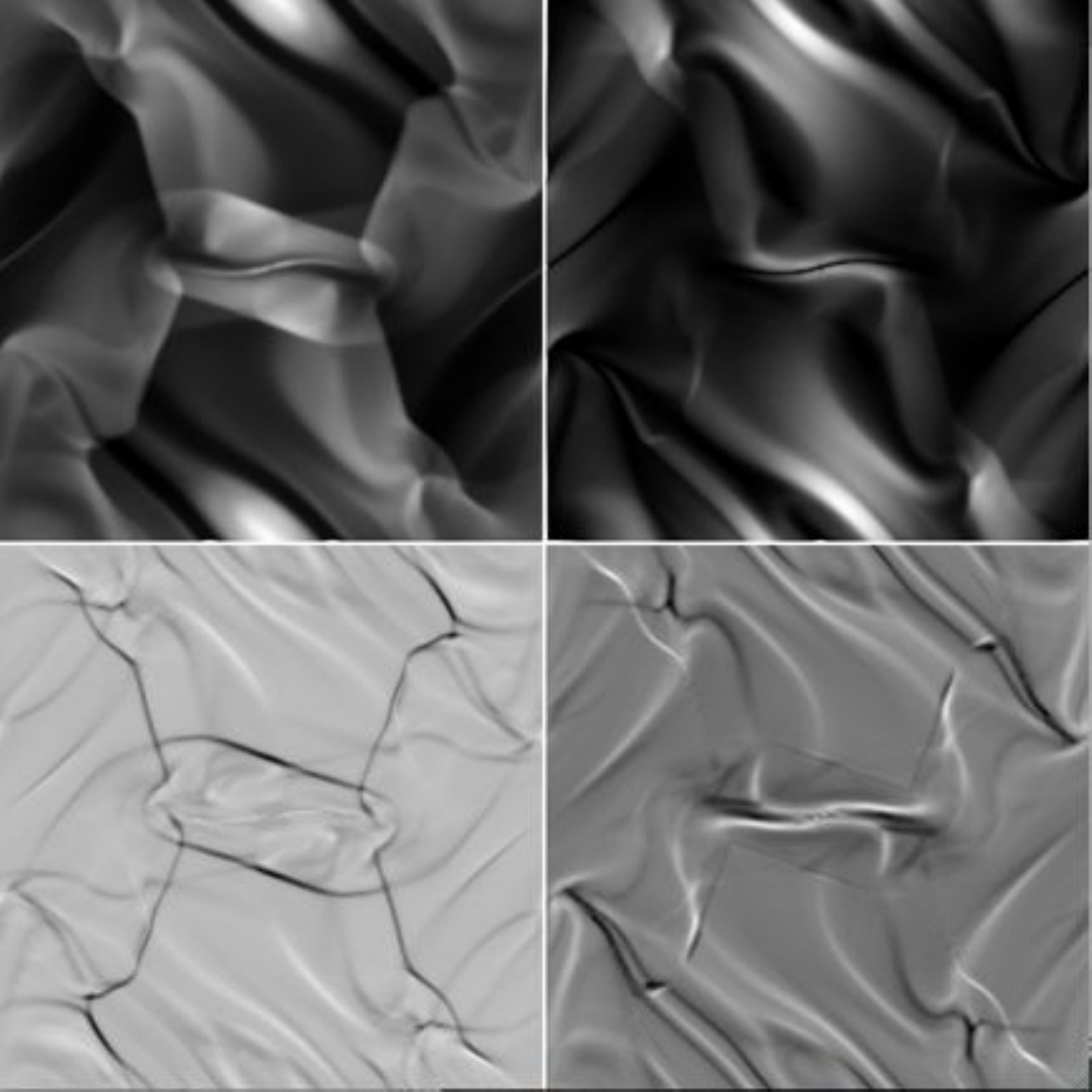}
 \caption{The result of the Orszag-Tang test on a $256\times 256$ grid, showing gas pressure (upper left), magnetic pressure (upper right), divergence of the velocity $\Div{\vec{u}}$ (lower left) and the rotation of the velocity $\left(\Rot{\vec{u}}\right)_{z}$ (lower right). This figure can be directly compared with fig. 3 of \citet{Ryu+etal98}.\label{fig:orsz}}
\end{figure*}

We have run a number of simulations with different grid resolution, diffusion and quench parameters. 
We observed good evolution of the shocks in all runs when we used
diffusion parameters set to $\nu_1>0.02$, $\nu_2>0.2$ and
$\nu_3>0.2$ in agreement with the previous tests. However, the shocks will be smoother when increasing the
diffusion or when decreasing the resolution. 
\Fig{fig:orsz} is created to be directly compared with Fig. 3 of \citet{Ryu+etal98} who use a upwind, total variation diminishing code, and even though there are small differences it is hard to tell which code is superior in spite of the two codes being fundamentally different.

%For example, in \Fig{fig:orsz} 
%the pressure is shown for the cases with $100\times 100$ and
%$200\times 200$ grid points and diffusion constants set 
%to $\nu_{1}=0.03$, $\nu_{2}=0.3$, $\nu_{3}=0.3$. 
%
%\begin{figure}
%  \resizebox{\hsize}{!}{\includegraphics{figs/orsz100.eps}}
%  \resizebox{\hsize}{!}{\includegraphics{figs/orsz200.eps}}
% \caption{\label{fig:orsz} The image shows the pressure  in a linear grey scale at time $0.633$ for examples with 100x100 grid points (top panel) and 200x200 grid points (bottom panel).}
%\end{figure}

\subsection{Test of chromospheric radiation approximation}

There are no analytical tests that can be used to test our recipes of the chromospheric radiation described
in \Sec{sec:genrad}. The best we can do is to make a comparison with a simulation where the 
approximated  processes have been calculated in detail. For this purpose we use 1D radiation
hydrodynamic simulations calculated with the RADYN code, see \eg\ Carlsson \& Stein (1995, 2002).
\nocite{1995ApJ...440L..29C} % Carlsson+Stein Radyn
\nocite{Carlsson+Stein02} % Carlsson & Stein 1D RADYN
This code solves the one-dimensional equations of mass,
momentum, energy, and charge conservation together with
the non-LTE radiative transfer and population rate equations,
implicitly on an adaptive mesh.

\begin{figure}
 \includegraphics[width=\columnwidth]{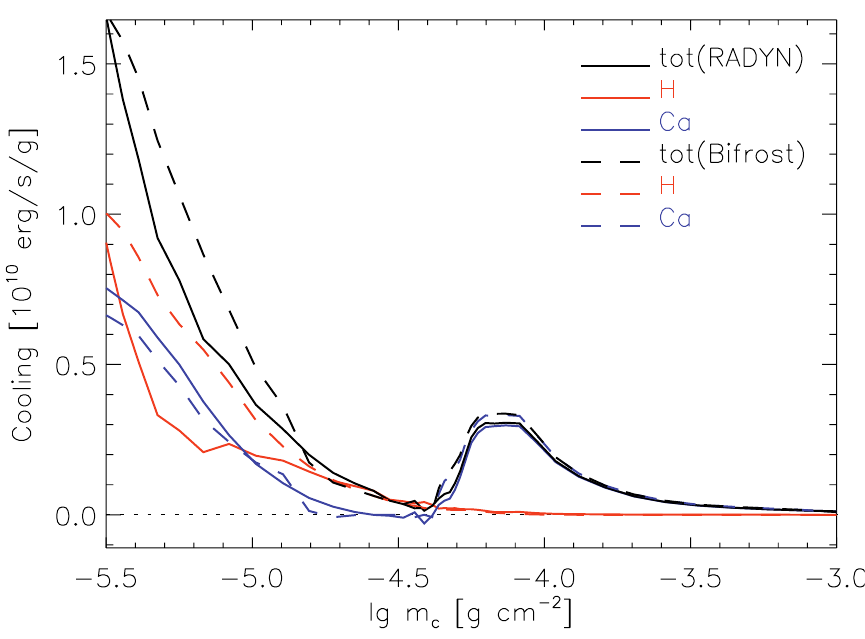}
  \caption{\label{fig:genradtest} Radiative cooling as function of column mass for a detailed 1D simulation with
  the RADYN code (solid) and with the \bifrost{} recipes for chromospheric radiation (dashed). Total cooling (black)
  and the contributions from hydrogen lines and the Lyman continuum (red) and lines from \ion{Ca}{II} (blue).}
 \end{figure}

Since we have used such a simulation to determine the free parameters in our recipes, we use a simulation 
with a different velocity field for the test. \Fig{fig:genradtest} shows the comparison of the 
radiative cooling for one timestep in the RADYN simulation with the cooling calculated with the 
\bifrost{} recipes. The timestep shown has a strong wave that is close to steepening into a shock at
lg($m_c$)=-4.2 (height of 1.2~Mm). The maximum temperature at the wave crest is 7000~K. Above the wave the temperature
rises rapidly into the corona. At lg($m_c$)=-5.5 the temperature is 9000 K. The \bifrost{} approximations
come close to describing the cooling in both hydrogen and calcium except that the cooling is
overestimated in hydrogen just below the transition region (left part of the figure). Inspection of the cooling as function of time at a given height shows that the recipe for hydrogen typically overestimates the cooling in the hot phases but does not include heating in the cool phases with the integral over time being close to the RADYN results. For
further tests of the chromospheric radiation approximations see Carlsson \& Leenaarts (in preparation).
%\citet{Carlsson+Leenaarts2011}.

\subsection{Full radiative transfer test in an isothermal scattering atmosphere}

\begin{figure}[htbp]
\includegraphics[width=\hsize]{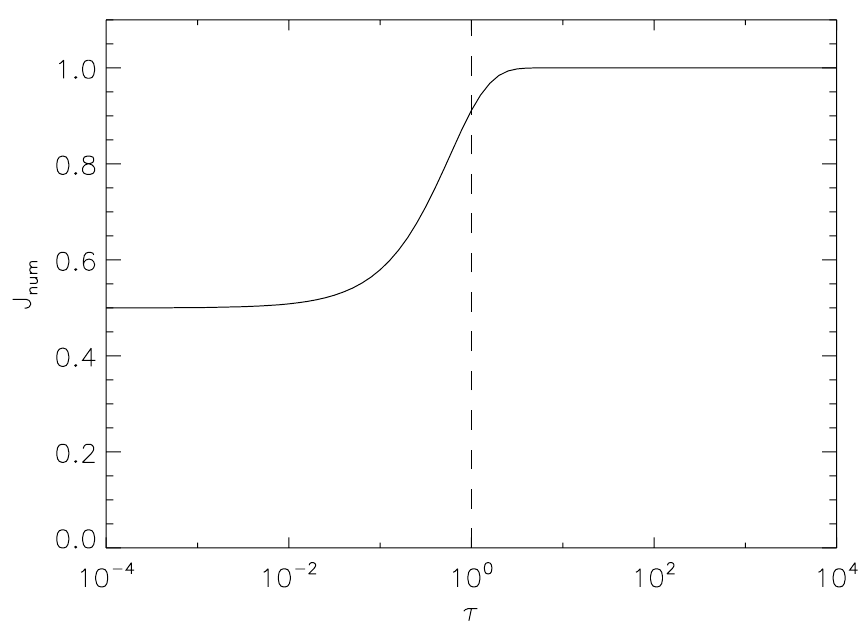}
\caption{Numerical solution for the mean intensity $J$ in LTE ($\epsilon=1.0$) as a function of optical depth $\tau$. The dashed line at $\tau=1$ marks the optical surface.}
\label{fig:jmeanlte}
\end{figure}

\begin{figure*}
\centering
\includegraphics[width=17cm]{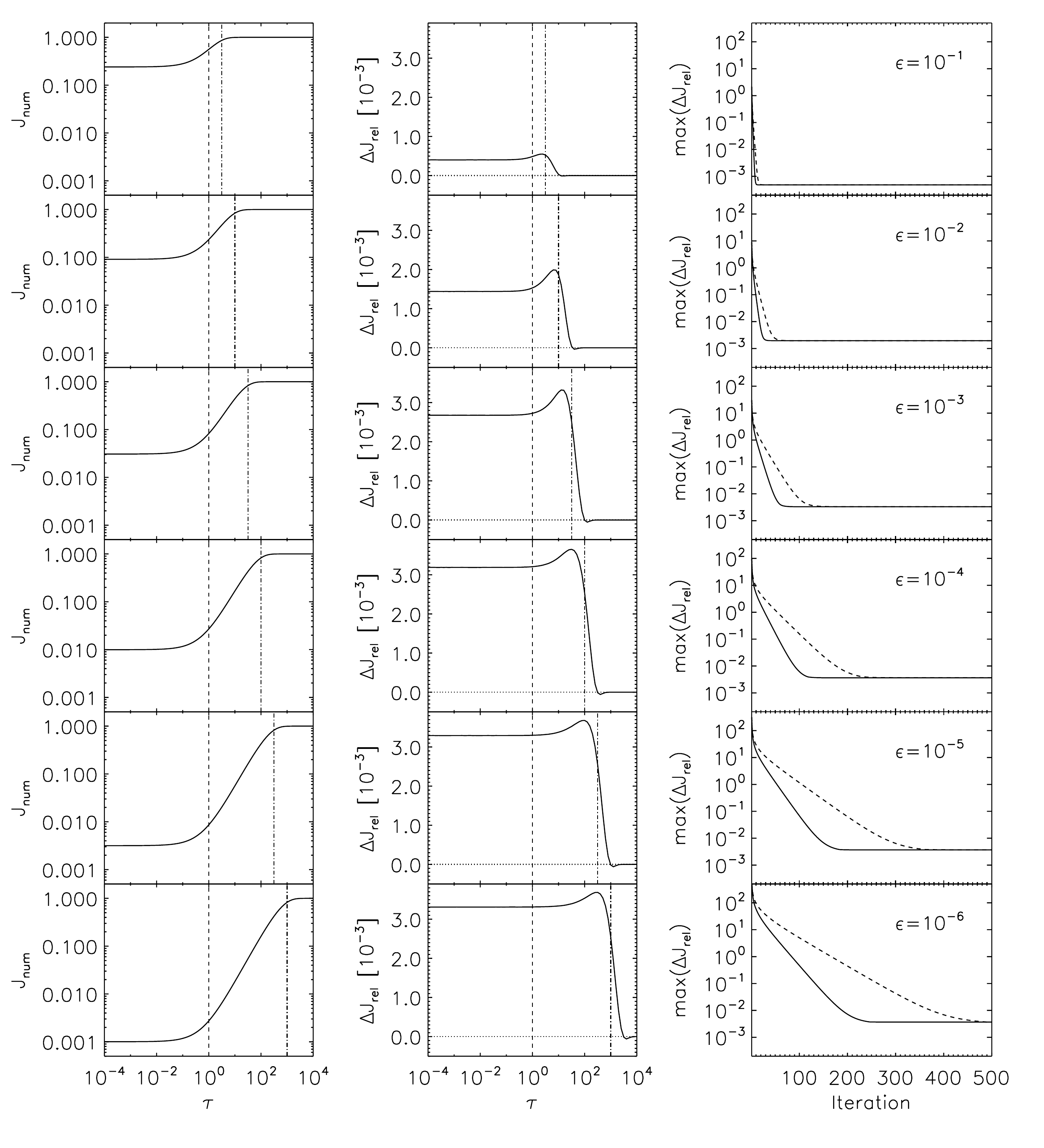}
\caption{Numerical solution for the mean intensity $J_{\mathrm{num}}$ as a function of optical depth $\tau$ (left column), relative deviation $\Delta J_{\mathrm{rel}}$ from the analytical solution as a function of optical depth $\tau$ (center column), and maximum relative deviation $\max(\Delta J_{\mathrm{rel}})$ as a function of iteration count (right column). The photon destruction probability $\epsilon$ ranges from $10^{-1}$ (top row) to $10^{-6}$ (bottom row). The dashed and dot-dashed lines in the left column and center column mark the optical surface at $\tau=1$ and the thermalization depth $\tau_{\mathrm{therm}}$. The dotted line in the center column indicates zero deviation. Dashed lines in the right column show the convergence speed for Gauss-Seidel corrections applied only during upsweeps, solid lines show the convergence speed for corrections applied during both upsweeps and downsweeps.}
\label{fig:jmeanconverge}
\end{figure*}

The radiative transfer equation with coherent scattering has analytical solutions in the case of a static 1D plane-parallel isothermal atmosphere if the photon destruction probability $\epsilon$ is constant at all depths \citep[e.g.,][]{RybickiLightman:1979}. The anisotropy of the radiation field needs to be restricted to linear dependence on the cosine of the zenith angle, $\mu=\cos\theta$, making the Eddington approximation exact and reducing the problem to solving a second-order equation for the mean intensity $J$. This setup is also known as the ``two-stream'' approximation. As second-order Gauss-Legendre quadrature yields an exact representation of integrals over a linear polynomial, a numerical result for $J$ becomes directly comparable to an analytical solution \citep[cf.][]{TrujilloBuenoetal:1995}.

We set the radiative transfer solver to reproduce the mean radiation field
\begin{equation}
J_\mathrm{an}(\tau)=\left[1-\frac{e^{-\sqrt{3\epsilon}\tau}}{1+\sqrt{\epsilon}}\right]B\label{eqn:jan}
\textrm{\ ,}
\end{equation}
with the source function
\begin{equation}
S(\tau)=\left[1-\left(1-\sqrt{\epsilon}\right)e^{-\sqrt{3\epsilon}\tau}\right]B\label{eqn:san}
\textrm{\ ,}
\end{equation}
where $\tau$ is the optical depth in the atmosphere, $\epsilon$ is the photon destruction probability and $B$ is the Planck function. The full radiative transfer module of the \bifrost{} code operates on a static 3D atmosphere with a resolution of $50\times50\times120$ grid points and with a horizontally homogeneous isothermal stratification; other physics modules are not used during the computation. The interpolation algorithms needed for radiative transfer in 3D simulations are validated separately with the searchlight test described in \citet{Hayeketal:2010}. Optical depths are preset on a grid that is equidistant in $\log_{10}\tau$ with 10 grid points per decade. Specific intensities are computed at $\mu=\pm1/\sqrt{3}$ and arbitrary azimuth angles $\phi$. The radiative transfer code uses double precision arithmetic to handle strong scattering at large optical depths, which appears due to the constancy of $\epsilon$ in the atmosphere. The Planck function is set to $B=1.0$ in arbitrary units for all computations; it is also used as a first-guess source function for the solver.

In the LTE case ($\epsilon=1$), the solver delivers $I^{+}(\tau)=1.0$ for outgoing intensities at $\mu=1/\sqrt{3}$ and $I^{-}(\tau)=1.0-e^{-\sqrt{3.0}\tau}$ for ingoing intensities at $\mu=-1/\sqrt{3}$ (see \Fig{fig:jmeanlte}). The numerical solution is equivalent to the analytical solution given by \Eq{eqn:jan}, as the Gauss-Seidel solver uses the formal solution of the radiative transfer equation, which leads to identical expressions. In the scattering case, photon destruction probabilities assume values between $\epsilon=10^{-1}$ (moderate scattering) and $\epsilon=10^{-6}$ (strong scattering), decreasing by factors of 10. The left column of \Fig{fig:jmeanconverge} shows the numerical results for $J$. The mean intensity near the surface decreases for smaller $\epsilon$ due to outward photon losses, and the thermalization depth $\tau_{\mathrm{therm}}=1/\sqrt{\epsilon}$, where the radiation field is completely thermalized ($J\approx B$), moves deeper into the atmosphere (dot-dashed line). At the smallest optical depths, the numerical solution delivers $J_{\mathrm{num}}(\tau=10^{-4})\approx\sqrt{\epsilon}$ for small $\epsilon$, as expected from \Eq{eqn:jan}.

The center column of \Fig{fig:jmeanconverge} shows the optical depth-dependence of the total error of the numerical solution as the relative deviation between $J_{\mathrm{num}}$ and $J_{\mathrm{an}}$,
\begin{equation}
\Delta J_{\mathrm{rel}}(\tau)=\frac{J_{\mathrm{num}}(\tau)-J_{\mathrm{an}}(\tau)}{J_{\mathrm{an}}(\tau)}
\textrm{\ .}
\label{eqn:jrelerror}
\end{equation}
At large optical depths $\tau\gg1$, radiative transfer is local and the total error depends on the source function gradient through the discretization error of the logarithmic optical depth grid. Using \Eq{eqn:san}, one obtains the variation $\Delta S$ of the source function between adjacent grid points with constant spacing $\Delta\log\tau$,

\begin{eqnarray}
\Delta S&\approx&\tau\frac{dS}{d\tau}\Delta\log\tau \nonumber\\
&=&(1-\sqrt{\epsilon})B\left(\sqrt{3\epsilon}\tau e^{-\sqrt{3\epsilon}\tau}\right)\Delta\log\tau\nonumber\\
&=&(1-\sqrt{\epsilon})B\left(\sqrt{3}\left(\tau/\tau_{\mathrm{therm}}\right)e^{-\sqrt{3}\tau/\tau_{\mathrm{therm}}}\right)\Delta\log\tau
\label{eqn:dsdlogtau}
\end{eqnarray}

In the LTE case, $\Delta S=0$ everywhere in the atmosphere since $\epsilon=1.0$ and $S=B=1.0$, independent of $\Delta\log\tau$, and the accuracy of the numerical solution does not depend on the grid resolution. For $\epsilon<1.0$ and at optical depths $\tau\gtrsim\tau_{\mathrm{therm}}$, the radiation field is entirely thermalized ($J_{\mathrm{num}}\approx B=1.0$), and $\Delta J_{\mathrm{rel}}$ vanishes since $\Delta S\rightarrow0.0$. The source function starts to decrease quickly through the decreasing mean intensity near $\tau_{\mathrm{therm}}$ due to outward photon losses, causing a sharp increase in the error of the numerical solution. $\Delta J_{\mathrm{rel}}$ peaks in the translucent zone as $\Delta S$ is largest between $1.0<\tau<\tau_{\mathrm{therm}}$. The magnitude of the peak grows with decreasing $\epsilon$ through the $(1-\sqrt{\epsilon})$ factor in \Eq{eqn:dsdlogtau}; an upper limit is reached with $\epsilon\rightarrow0.0$. It also follows from \Eq{eqn:dsdlogtau} that $\max(\Delta J_{\mathrm{rel}})$ scales with the grid resolution. Further up in the atmosphere, the error decreases again through the finer grid spacing. At optical depths $\tau\ll\tau_{\mathrm{therm}}$, the local discretization error becomes negligible due to the vanishing $\Delta S\rightarrow0.0$. However, as $I^{+}$ and $I^{-}$ decouple from the local source function and the radiation field becomes anisotropic, the error becomes independent from the source function gradient. $\Delta J_{\mathrm{rel}}$ is dominated by the propagated error of outgoing radiation $I^{+}$ from deeper layers and is therefore constant. 

The right column of \Fig{fig:jmeanconverge} shows the convergence speed of the numerical result to the analytical solution, measured through the maximum relative deviation $\max(\Delta J_{\mathrm{rel}})$ in the atmosphere (see \Eq{eqn:jrelerror}). Dashed lines represent computations where Gauss-Seidel corrections for the source function were applied only during upsweeps, while solid lines show the results when both upsweeps and downsweeps were corrected, which increases convergence speeds by about a factor of two \citep[see][]{TrujilloBuenoetal:1995}. Convergence is significantly slower when the thermalization depth moves to deep, very optically thick layers for small photon destruction probabilities, requiring about 250 iterations in the strongest scattering case. We find a numerical error of $\max(\Delta J_{\mathrm{rel}})\approx3.7\cdot10^{-3}$ at $\epsilon=10^{-6}$, similar to the error quoted in \citet{TrujilloBuenoetal:1995}.

\subsection{Hydrogen ionization test}

 \begin{figure}
 \includegraphics[width=\columnwidth]{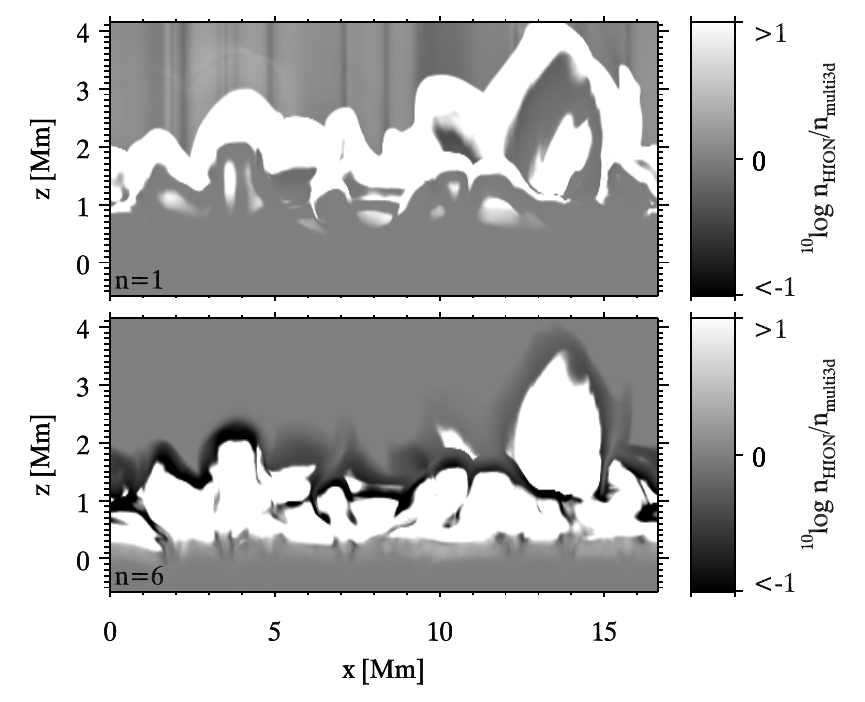}
  \caption{\label{fig:hiontest} Ratio of the hydrogen level
    populations computed with the hydrogen ionization module
    to the populations from a detailed computation assuming
    statistical equilibrium. Upper panel: atomic hydrogen ground level;
    lower panel: proton density.}
 \end{figure}

Non-equilibrium hydrogen ionization generally produces atomic level
populations that are neither in statistical equilibrium nor in
Saha-Boltzmann equilibrium (LTE). However, in the case of simulations of the solar atmosphere
spanning from the upper convection zone to the lower corona, two
limiting cases are produced: near the convection zone boundary the
level populations obey LTE statistics; near the
coronal boundary the populations obey statistical equilibrium
because of the fast transition rates there. Both limiting cases are
reproduced in statistical equilibrium 
non-LTE radiative transfer codes. 
%Mats: added the following to make it extremely clear that it is only
%in the limits the test make any sense at all
We will therefore compare the results from a \bifrost{} run with those
of a statistical equilibrium code {\em in those two limits}. In the
intermediate regime the statistical equilibrium is not valid and there
will be large differences between the two codes.

We  took a snapshot from a 2D simulation of the solar atmosphere that included the hydrogen ionization module
(HION). This simulation had a grid size of $512 \times 325$ points,
spanning from 1.5~Mm below the photosphere to 14~Mm above it. It
included a weak magnetic field (average unsigned flux density in the
photosphere of 0.3~G), thermal conduction and full, chromospheric and
optically thin radiative transfer. The convection zone boundary was
open, the coronal boundary used the methods of characteristics.

We then computed the statistical equilibrium
values of the hydrogen level populations from this snapshot using the
radiative transfer code Multi3D treating each column as a 1D
plane-parallel atmosphere. As input we took the snapshot geometry, the
mass density, the electron density and the temperature. We set the
velocity field to zero. As model atom we used a 5-level plus continuum
hydrogen atom with all radiative transitions from the ground level put
in radiative equilibrium, as is assumed in the HION module. We
treated all remaining lines assuming complete redistribution.

\Fig{fig:hiontest} shows a comparison of the level populations
obtained from the \bifrost{} and the Multi3D computation. Both the $n=1$
and the proton density are equal in the convection zone (below
$z=0$~Mm), showing that \bifrost{} correctly reproduces LTE populations
there. The proton densities in the corona are also equal for \bifrost{}
and Multi3D. The $n=1$ populations in the corona are not identical.

However, the differences are small, the populations are always within a
factor of two of each other. The exact value of the coronal population
density depends on the radiation field, and the simple HION assumption
of a coronal radiation field that is constant in space does not
capture the small-scale variations present in the Multi3D computation
as indicated by the vertical stripes in the upper panel of
\Fig{fig:hiontest}. This striping is caused by variations in the
atmospheric quantities down in the photosphere and chromosphere where
the coronal radiation field is set.

\subsection{Boundary condition test}
\label{sec:bc_test}

The full 3D MHD equations allow a wide variety of wave modes that
cross the boundaries at different angles depending on their origin and
on the topology and strength of the magnetic field. A comprehensive
test of all these modes falls outside the scope of this paper, but we
will present a couple of tests to show the range of boundary condition
behavior we have observed. The examples shown contain the full set of
physics that \bifrost{} supports and thus are meant to represent the
`typical' production run the code is meant for, though with simpler
geometries. Both of the tests shown were at the same time used to test
the thermal conduction module as described in \Sec{sec:conduction_test}.

\begin{figure}
  \resizebox{\hsize}{!}{\includegraphics{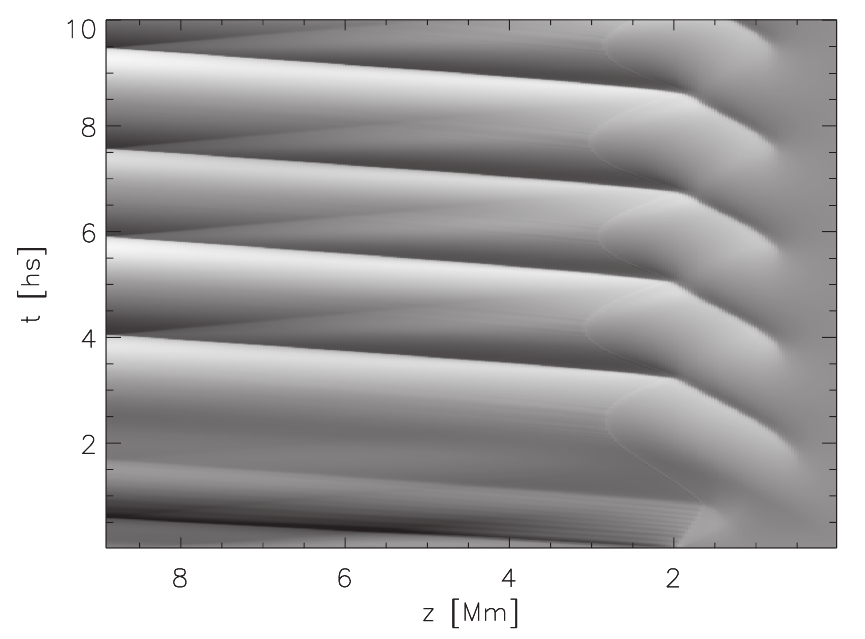}}
 \caption{\label{fig:bc_uz_1d1} Vertical velocity $u_z$ in 1D test
   model. The color scale is set to span $\pm 30$~km/s with black
   representing upflow and white downflow. The chromosphere adjusts its
   structure, which initially is not quite in hydrostatic equilibrium,
   by emitting acoustic waves at the cut-off frequency. These waves
   are initially strong enough to form shocks, as here during the
   first 1000~s of the simulation.
 }
\end{figure}

The first test is a 1D model containing a photosphere, chromosphere,
and corona that has a vertical extent $z=9$~Mm, which is discretized on
a grid containing 256 points with $\Delta z=35$~km throughout the
computational domain. 
The model has solar
gravity and a weak ($0.1$~G) vertical magnetic field. Radiative losses
are included through the optically thin and chromospheric
approximations, but the full radiative transfer module is turned
off. Thermal conduction is included and the corona is kept heated
by maintaining the temperature at the upper boundary at 1.1~MK. The
initial model is not perfectly in hydrostatic nor energetic equilibrium
and the atmosphere responds by launching a set of acoustic
disturbances at roughly the acoustic cut off frequency of 3~minutes.
These are initially of high amplitude, forming shock waves, but at
later times damping out to much lower amplitudes and forming linear
waves. In \Fig{fig:bc_uz_1d1} we plot the vertical velocity
$u_z$ as a function of height and time. Evidently, acoustic waves
originate in the lower to middle chromosphere near $z\approx 500$~km
and propagate upward steepening into shocks. At roughly 2~Mm they
encounter the transition region before entering the corona proper. 

It is clear that while the boundary seems fairly well behaved, the 
strongest shocks do lead to some reflection, which we have measured to
be of order 5\% or less in terms of reflected energy. There may be %are? 
several reasons for such reflection: One is that high amplitude waves 
seem in general more difficult to transmit through a boundary.
In addition, setting a hot plate as a temperature boundary condition
will give some reflection as the temperature is forced to a given
value. 

\begin{figure}
  \resizebox{\hsize}{!}{\includegraphics{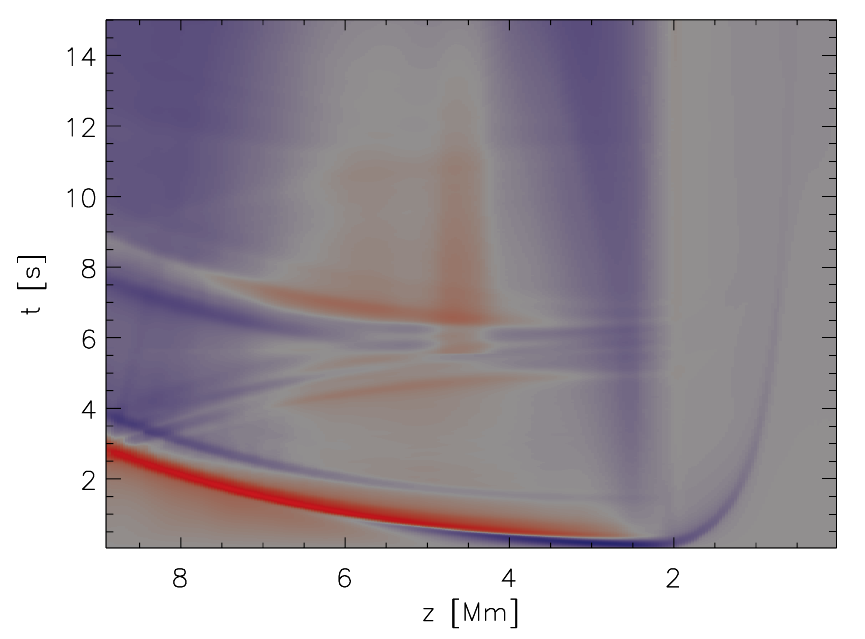}}
 \caption{\label{fig:bc_uz_2d80} Vertical velocity $u_z$ in 2D test
   model, red is downflow and blue upflow with a color scale set to
   $\pm 50$~km/s, as a function of time and height at the position
   $x=5$~Mm in our 2D test model (see text). The disturbance clearly visible
   is a fast mode wave that originates due to a slight imbalance in
   the Lorentz force in the transition region in the initial
   state. The wave propagates fairly cleanly through the upper
   boundary, with a reflection of some 5\%.  
}
\end{figure}

The second experiment uses the same background atmosphere as in the 1D
test described above, but we have expanded to span $16.5$~Mm in the
horizontal direction forming a 2D atmosphere that contains a
magnetic field of greater complexity: This magnetic field is formed by inserting
positive and negative magnetic polarities of $\pm 1000$~G strength at the
bottom boundary, spanning 1~Mm and centered at $x=2.5$~Mm and $x=13.5$~Mm respectively, 
and then computing the potential field that arises from this distribution (see
\Fig{fig:fc_2d}). Note that the field is quite strong and that
plasma beta is less than one in most of the modeled domain. Again
there is a slight hydrostatic imbalance in the  initial state
and a transient wave is generated, in this case in the form of a fast
mode wave originating close to the transition region. The
Alfv{\'e}n speed is quite high, some $9\,000$~km/s near the
transition region, falling to $1\,000$~km/s at the upper boundary, since much of the field has closed at lower heights in
the atmosphere. In comparison, the speed of sound lies in the range
$100$~km/s to $160$~km/s in the corona.
Thus, the generated fast mode, traveling essentially at the Alfv{\'e}n
speed, propagates to the upper
boundary very quickly, using only 3~s to cover the 7~Mm from the
transition region to the upper boundary. The transient wave's amplitude on leaving the upper boundary
depends on location but lies between $-70$~km/s and $150$~km/s.
In \Fig{fig:bc_uz_2d80} we plot the vertical
velocity as a function of height and time at the representative
horizontal position $x=5$~Mm. As in the previous example we do
find a certain amount of reflection, but again find it to be energetically
unimportant, at less than 5\% of the energy contained in the original
wave at all horizontal locations. 
Note that the reflected wave is re-reflected off the transition region 
and that this second wave seems very nicely transmitted through the 
upper boundary.

Setting boundary conditions for the type of models discussed here is a
compromise between long term stability and the best possible transmission
of outwardly propagating modes. In the examples presented above we
have attempted to show the results where stability is the most
important factor, which is attained by limiting the velocity amplitude
allowed in the ghost zones, as would be typical for a production run
that aims to model the outer solar atmosphere for a duration of order
several thousand seconds. This safety factor increases the amount of
reflection seen, a compromise that allows us to run simulations for
long periods of time.

\subsection{Thermal conduction test}
\label{sec:conduction_test}

\begin{figure}
  \resizebox{\hsize}{!}{\includegraphics{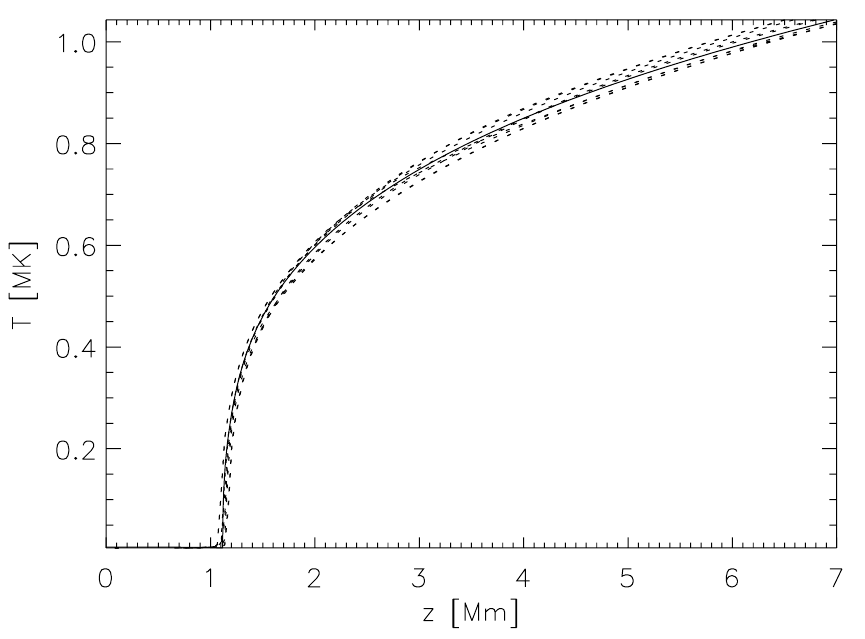}}
 \caption{\label{fig:fc_1d} Comparison of high resolution 1D model
   (solid line $\Delta z\gtrsim 80$~m) with \bifrost{} test run with $\Delta
   z=35$~km (dashed lines). The models are set up to have a
   temperature maximum of $1.1$~Mm at the upper boundary such that
   conduction dominates the energetics of the atmosphere in the corona
   and transition region. The results from the \bifrost{} run are
   taken from the same model as used in \Fig{fig:bc_uz_1d1},
   but at much later times ($t>8\,000$~s) when acoustic perturbations 
   are largely damped and the amplitude of transition region motion is
   less than $300$~km.
}
\end{figure}

In order to test the thermal conduction module we present the
temperature structure that arises in the two examples presented above
in \Sec{sec:bc_test}. In the first we
consider a 1D model with vertical magnetic field of the upper
photosphere, chromosphere and corona in which the chromosphere is slowly
relaxing by shedding acoustic waves at the cut-off frequency. The
upper boundary temperature is set to 1.1~MK and conduction is the
dominant term in the energy balance in the corona and transition region,
down to temperatures of roughly 10~kK occurring at a height of
$z=1.2$~Mm. After 3~hours solar time the amplitude of the
acoustic waves being generated is much reduced, and the location of 
the transition region moves by less than 300~km during a full wave
period. In \Fig{fig:fc_1d} we plot the temperature as a
function of height for several timesteps during a wave period 3~hours
solar time after the experiment began. Also plotted is the
temperature profile that arises from a much higher resolution 1D model
computed with the TTRANZ code \citep{1993ApJ...402..741H} with the
same upper boundary temperature. The latter 
model uses an adaptive grid \citep{1987JCoPh..69..175D} that concentrates 
grid points in regions of strong gradients; in the present model the
grid size is of order 80~m in the lower transition region. The 
temperature profiles in the two models are quite similar even though
the grid spacing is much larger in the \bifrost{} run. Of course, not all
aspects of the relevant physics can be reproduced with $\Delta
z=35$~km in the high temperature gradient environment of the lower
transition region. We find sudden large temperature changes
in transition region grid points as the location of the transition
region changes due to the passage of acoustic waves. These temperature
changes become sources of (higher frequency) acoustic waves with
amplitudes of order some 100~m/s that can be discerned on close
inspection of \Fig{fig:bc_uz_1d1}. This artifact first
disappears when the grid size is set to $\Delta z \lesssim 15$~km (for
typical coronal temperatures $\lesssim 2$~MK).

In the second example we consider the case of rapid heating of coronal
plasma in a magnetized atmosphere and follow how thermal conduction
leads the deposited energy along the magnetic field lines. We use the
same atmosphere as described in the 2D case given in the boundary
condition test above. During the first second of the model run we
deposit $50$~J/m$^3$ over a region spanning $100\times 100$~km$^2$ (an
amount of energy equivalent to a large nano-flare) at location
$x=7$~Mm, $z=6.3$~Mm, after which we allow the atmosphere to cool. At
the upper boundary we set a zero temperature gradient boundary
condition. 

\begin{figure}
  \resizebox{\hsize}{!}{\includegraphics{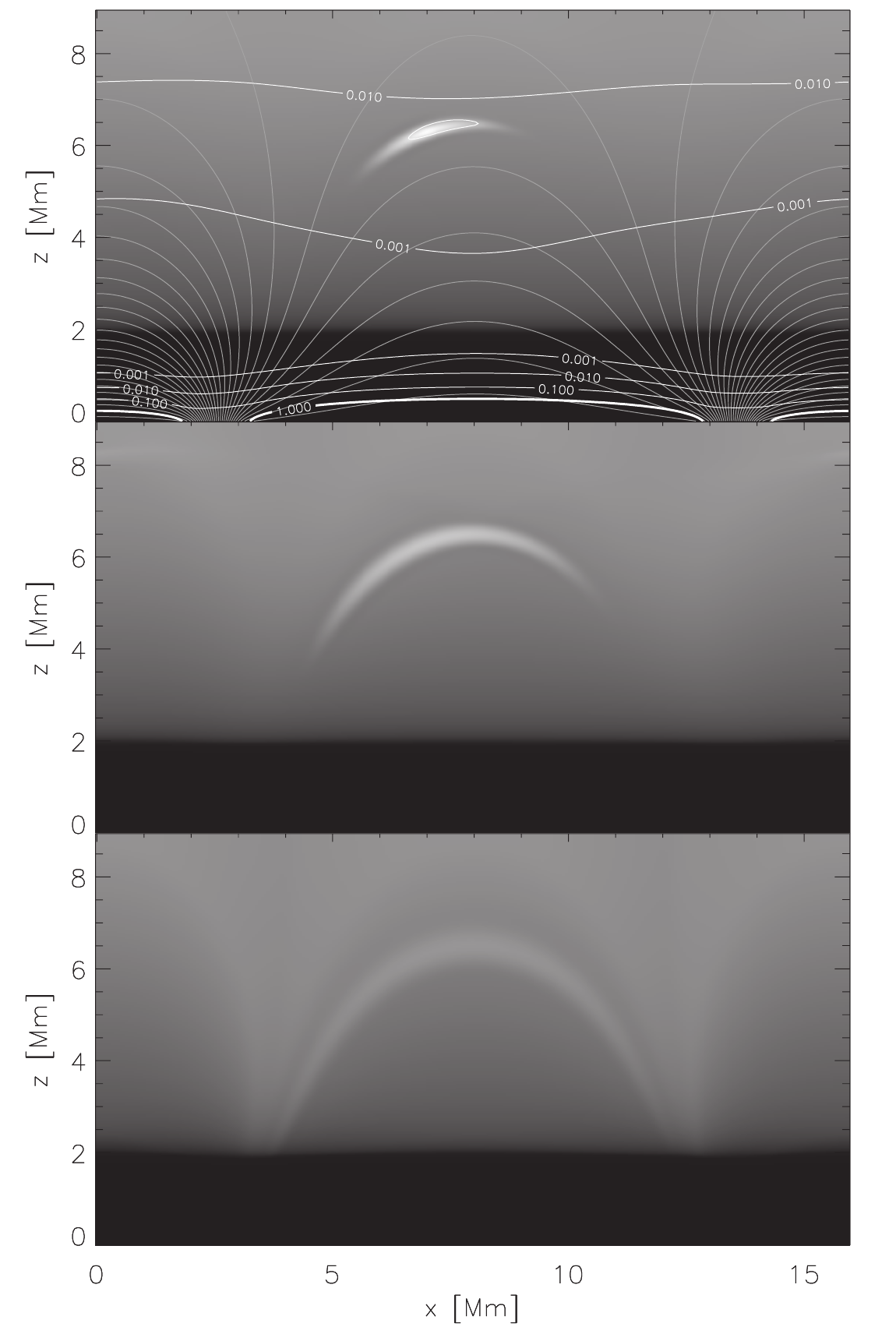}}
 \caption{\label{fig:fc_2d} Time evolution of 2D model which is heated
 for 1~s at position $x=7$~Mm, $z=6.3$~Mm with 50~J/m$^3$ over a
 region spanning $100\times 100$~km$^2$.  The panels show the
 temperature at $t=0.6$~s (top) when the maximum temperature is
 $2.25$~MK, at$t=1.8$~s (middle) when the maximum temperature is
 $1.63$~MK, and at $t=10$~s (bottom) where the maximum temperature is 
$1.16$~MK. The temperature increases rapidly in the heated region,
 reaching 2.4~MK at $1$~s, and plasma is heated by thermal conduction
 along the field as the plasma cools. Note that the upper boundary is
 set to have zero temperature gradient, so the entire atmosphere
 cools as well. Magnetic field lines are indicated with thin grey contours
 and contours of constant $\beta$ are shown with white numbered lines.
}
\end{figure}

\Fig{fig:fc_2d} shows the temporal evolution of this model with
snapshots taken at $0.6$~s, $1.8$~s, and $10$~s. The deposited energy
rapidly heats the coronal plasma, originally at 1~MK, to $2.4$~MK at
$t=1$~s. The heat is efficiently conducted away from the site of energy
deposition and has already after $0.6$~s increased temperatures in a
region several thousand kilometers along the magnetic field. After
energy deposition ends the maximum temperature in the heated region
decreases while the region itself continues to expand along the
magnetic field towards the transition region and chromosphere. At $t=10$~s
the heated region has cooled quite a bit, but is still hotter than the 
ambient atmosphere and has spread out to form a loop-like structure. 
Note that the ambient atmosphere is also
cooling, the portions of the atmosphere connected via the magnetic
field to the upper boundary cooling least rapidly. The width of the
heated region is fairly constant, but some spreading perpendicular to
the magnetic field is evident in the last frame shown at
$t=10$~s. 

\section{Parallelization}
\bifrost{} was written to be massively parallel. 
It employs the Message Passing Interface (MPI), because it is very well developed and exists on almost all super computers. There are a number of other options for parallelization, including OpenMP and the {\texttt{thread()}} mechanism included in the different variations of the C programming language, but we found MPI to suit our needs the best. 

MHD on a regular grid is trivial to parallelize by splitting the computational grid into subdomains and distribute one subdomain to each node. In the case of \bifrost{}, computing the derivatives or interpolating the variables uses a stencil of 6 grid points. The two/three gridpoints nearest the edge of a subdomain, then relies on data that is outside the subdomain belonging to the local node, but instead belongs to the neighboring subdomain. The way this data is acquired most efficiently, depends on the problem. The two possible solutions are to communicate the needed data to neighboring nodes every time it is needed, and the other is to supply each node with a number of ``ghost cells'' around its allocated subdomain so that the whole stencil used in for instance a derivative operator is present at the node that needs it. The extreme solution would be for each node to have a very large number of ghost cells, in the most extreme case the complete computational grid, and therefore never need to communicate, but that would make the code non-parallel. Choosing to include ghost cells around each sub domain makes it necessary to do more computations, since the ghost cells are copies of cells belonging to neighboring nodes, but makes it possible to do less communication, as the ghost cells can be filled with the correct values from the neighboring nodes less often. The number of ghost cells chosen is therefore influenced by the relative importance of the communication speed, the computation speed and the numerical scheme of the code.

\begin{figure}
  \resizebox{\hsize}{!}{\includegraphics{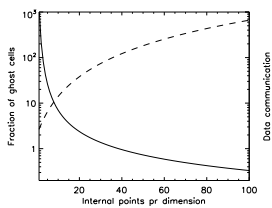}}
  \caption{The relative number of computed gridpoints (solid) and the total data needed to be communicated (dashed) as a function of the number of internal grid points per dimension for a 3D cubic sub domain}
   \label{fig:comp_vs_comm}
\end{figure}

We have chosen to keep the communication between nodes to a minimum at the expense of doing more computations. Five ghost cells makes it possible to do both a derivative and an interpolation along the same coordinate direction without having to communicate with neighboring nodes. That choice was made, because it allows us to get good performance even on machines with relatively slow internode communication speeds. It is very hard to quantify exactly when this choice is wise and when not, because so many parameters enter the problem. For instance, a global minimum operation scales  with the number of nodes used and the communication speed, a simple neighbor communication will depend both on the communication time, the physical setup of the nodes and the switches that connect them etc. But it is worth noting that typical communication times in large systems are of millisecond scale, while the frequency of the cpus are in the gigahertz range. So from this very simple argument, it should be possible for a CPU to do roughly $10^3$ computations (multiplications, additions), in the time it takes to do one communication. 

\Fig{fig:comp_vs_comm} shows how the data communicated increases with the number of internal gridpoints per node, and how the relative increase in computation decreases with the number of internal grid points. Both of the curves in \Fig{fig:comp_vs_comm} do not take into account communication speed or computation speed, so both curves can be shifted up and down the y-axis when applying them to a specific system. Communication can be split into initialization and actual communication of the data. In general the initialization takes very long compared with the actual transmission of the data, so there is a large offset in the communication time, but the MHD part of \bifrost{} only communicates with neighboring subdomains, so this offset depends on the dimensionality of the problem, and not on the number of cpus or internal points per subdomain. Since \bifrost{} was developed to handle as large computations as possible, the number of internal grid points will be rather high, so the rapid divergence between the two curves makes it plausible that doing the extra computations is the correct choice. 

Scaling can be measured in two different ways, called strong and weak scaling. Strong scaling uses a set problem size and then timing measurements are made for different number of nodes, while weak scaling uses a set problem size per node, which means that when the number of nodes increases, the problem size also increases. Strong scaling is mainly a test of the communication overhead. If the communication takes up a constant amount of time, it should take a relatively larger and larger part of the run time as the number of cpus goes up. Weak scaling gives a measure of how well the code handles larger and larger problem sizes and number of cpus without being influenced by the raw communication time. 

\begin{figure}
  \resizebox{\hsize}{!}{\includegraphics{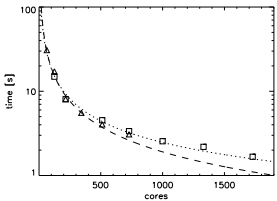}}
  \caption{Scaling when running a pure MHD case with $500^{3}$ gridpoints on a Cray XT4 system with different number of cores (triangles), the theoretical scaling curve (dashed) and the theoretical scaling curve taking into account ghost cells (dotted)}
   \label{fig:strong-scaling}
\end{figure}

Both strong and weak scaling tests of \bifrost{} have been performed on a number of
computer architectures. We will here report on the results from a Cray XT4, with each node containing one quad-core AMD Opteron 2.3 GHz cpu and with a proprietary Cray Seastar2 interconnect, made available to us by the Partnership for Advanced Computing in Europe (PRACE), and from a Silicon Graphics ICE system, with each node containing two quad-core Intel Nehalem-EP 2.93 GHz cpus with Infiniband interconnect, located at NASA Advanced Supercomputing Division. 

Strong scaling tests were run with just the simplest configuration: pure MHD
on an uneven staggered mesh, with an ideal EOS, without
radiation, conduction or any other advanced physics or boundary
conditions. The timing is performed on the Cray XT4 in such a way that the number of
internal points per dimension is 30 or more, so the lower end of
\Fig{fig:comp_vs_comm} is never reached. Such a test will show if
there are communication bottlenecks in the code which will
significantly slow the code down when running on a large number of nodes and when there is too large a penalty due to the use of ghost cells. 
\Fig{fig:strong-scaling} shows the scaling results for a $500^{3}$ gridpoint run, and as the number of cores increases the relative amount of grid cells that are ghost cells increases, and consequently the code uses more and more time calculating the ghost cells compared to the internal cells. If the relative increase in ghost cells is taken into account, \bifrost{} scales very well. \Fig{fig:strong-scaling} also shows that if the number of cells per dimension for each core becomes less than 50, then efficiency of the code has dropped by about 35\% compared to perfect scaling, and consequently, each core should not get a computational sub domain which is smaller than 50 grid points on a side. 

\begin{figure}
  \resizebox{\hsize}{!}{\includegraphics{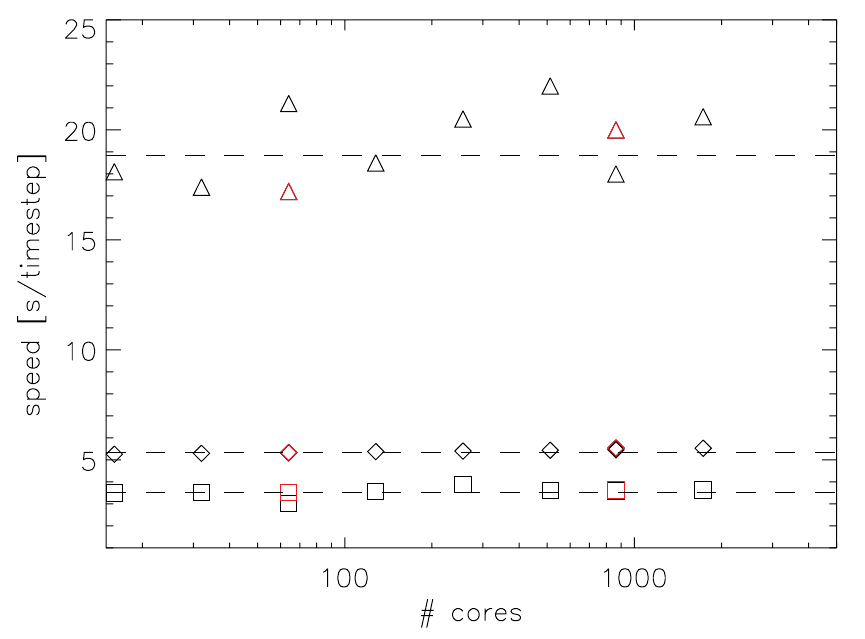}}
  \caption{Weak scaling results for \bifrost{} on a Cray XT4 system when running MHD with a realistic EOS and chromospheric radiation (squares), MHD with a realistic EOS, chromospheric radiation and Spitzer heat conduction (diamonds) and MHD with a realistic EOS, chromospheric radiation, Spitzer heat conduction and full radiative transport (triangles). Dashed lines show the average timing on the runs up to 256 cores.
}
  \label{fig:weak-scaling}
\end{figure}
\begin{figure}
  \resizebox{\hsize}{!}{\includegraphics{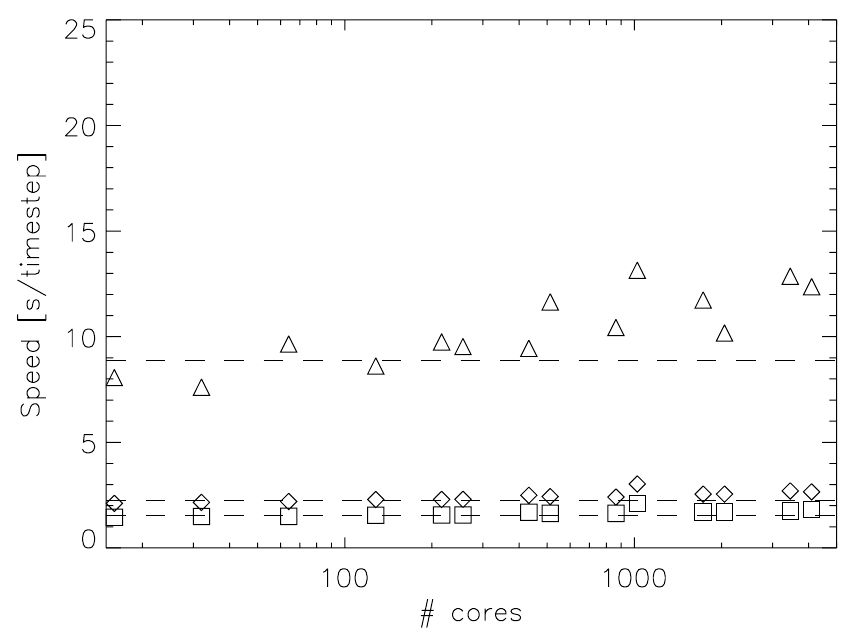}}
  \caption{Weak scaling results for \bifrost{} on a Silicon Graphics ICE system when running MHD with a realistic EOS and chromospheric radiation (squares), MHD with a realistic EOS, chromospheric radiation and Spitzer heat conduction (diamonds) and MHD with a realistic EOS, chromospheric radiation, Spitzer heat conduction and full radiative transport (triangles). Dashed lines show the average timing on the runs up to 256 cores.
}
  \label{fig:pleiades-scaling}
\end{figure}

The weak scaling tests were performed with a production setup of a solar simulation extending from the convection zone 2.5~Mm below the photosphere to the corona 14.5~Mm above the photosphere. The horizontal extent was $24\times 24$~Mm$^2$. When the full radiation module was switched on in some of the weak scaling tests, scattering was included and the radiation field was described with 26 rays. The bottom boundary was transparent, the top boundary used the method of characteristics and horizontal boundaries were periodic.

In theory \bifrost{} should be able to handle weak scaling very well, when only modules that use local data are used. If modules like full radiative transfer and Spitzer conductivity are included, the scaling is expected to drop below the theoretically best scaling, because these modules use non-local data and provide non-local results. When including modules using non-local data, the drop in efficiency with larger number of cpus depends on the choice of, and implementation of, the solver as well as communication time. For our tests the problem size was adjusted such that each core had a subdomain of $64\times64\times64$ internal grid-points. The weak scaling results for \bifrost{} on the Cray XT4 are provided in \Fig{fig:weak-scaling} and show very little dependence on the number of cores for each type of experiment. For the full radiative transfer case it is important to note that the number of iterations for convergence of the scattering radiative transfer problem depends on the change from the previous timestep, the aspect ratio between spacing in the vertical and horizontal directions and the resolution. The large fluctuation in the timings for the full radiative transfer case seen in \Fig{fig:weak-scaling} is completely caused by the variations in the number of iterations per timestep for the different cases and when this is taken into account the time per timestep does not increase with increased number of cores. \Fig{fig:pleiades-scaling} shows the same experiments but run on the Silicon Graphics ICE system with Infiniband interconnect. There is again little dependence on the number of cores when the trend of increasing number of iterations with increased resolution for the radiation is taken into account. 

\Figs{fig:weak-scaling}{fig:pleiades-scaling} also show that both the Spitzer heat conduction and full radiative transfer take up a large fraction of the time. Spitzer heat conduction increases the time by almost \mbox{50\%}, while full radiative transfer increases the computing time by more than a factor of 5. The large computational effort needed for the full radiative transfer is caused by the scattering iterations --- typically 3-15 iterations are needed for convergence within one timestep. In practical problems, the full radiative transfer does not have as large a penalty on the timing as it might seem, since the radiative transfer module only needs to be run in a fraction of the timesteps for chromospheric/coronal problems where scattering is important. In a typical solar simulation with 32~km horizontal resolution (the 1728 core points in \Figs{fig:weak-scaling}{fig:pleiades-scaling}) the timestep set by the Courant-Friedrich-Levy condition and the Alfv\'en speed in the corona is 3~ms while the radiation is updated only every 300~ms which means that the radiative transfer increases the computing time only by 2\%.
For photospheric problems, the radiative timescale is comparable with the hydrodynamic one and the radiative transfer module needs to be called for every timestep. On the other hand, there is no need to iterate when scattering is unimportant. For such simulations the radiative transfer increases the computing time by about 60\% compared with the pure MHD case.

\section{Conclusion}

The development of numerical methods and computer power has made numerical simulations of stellar atmospheres highly relevant. It is now possible to create observational predictions from advanced realistic numerical simulations, and observations can therefore partially validate the results of the numerical simulations. If such predictions are made and confirmed by observations it is likely that other predictions from the numerical simulations are also correct, making it possible to get much more information about the stellar atmospheres than would be possible through observations alone. To provide as good simulation results as possible, it is necessary to have a highly efficient and parallel numerical code. The tendency for modern super computers to be distributed memory systems, makes it extremely important that numerical codes are highly parallel. For pure MHD that is not difficult to attain because MHD is a local process, but it becomes much more complicated for non-local processes or numerical solvers. 

The numerical MHD code \bifrost{} has been created to meet the requirements of present supercomputers. It is developed by a group of researchers, post docs and PhD students and provides a simple interface to include further developments in boundary conditions and physical regimes that are not included in the simplified core of the code. Several extension modules are already provided and several more are under development. Results using \bifrost{} have already been published \citep{
          Hayeketal:2010,
          2010MmSAI..81..582C,   % Mats fra Sac Peak
          2010ApJ...718.1070H,    % Viggo om Red-shifts....
          Leenaarts++2011}. 
The pure MHD module of \bifrost{} has been extensively tested through standard tests and has performed well. There are at the moment eight individual modules that have been finished and tested, and several more are under development and testing. The finished modules include a number of equations of state, radiation transport and thermal conduction and tests of them have been presented and all produce results according to expectations.  

\bifrost{} has been tested for scaling performance. Both strong scaling and weak scaling results are very good on the two systems we have tested on. We would ideally have liked to test on a system where the interconnect is slower than on the Silicon Graphics ICE and Cray XT4 systems to  get further knowledge about the bottlenecks a slow interconnect would present for \bifrost{}. 
The hardware communication architecture plays a role on the scaling behavior of \bifrost{}, but these would most likely be more severe if we had made a choice of doing more communication. Since \bifrost{} is primarily designed to run large simulations, using a large number of cores, we believe we have made the correct prioritization in choosing more computations over communication.

The very good scaling performance and the modules already developed will make it possible to simulate the whole solar atmosphere from the top of the convection zone to the corona with a degree of realism that has not been attained before. It has become more and more clear that the solar atmosphere cannot be split into the traditional separate layers, the photosphere, chromosphere, transition region and corona. The solar atmosphere is one large connected system, and it is necessary to include the whole atmosphere to attain credible results. The consequence is that the code used for such a simulation will have to include the special physics important in each layer, making the numerical code much more complex than a numerical code designed to deal with just one of the layers. \bifrost{} is uniquely qualified for that task. 

The relative ease of creating new setups for simulations, inclusion of special physics and boundary conditions, make it possible to use \bifrost{}  for detailed solar atmosphere simulations and for stellar atmospheres in general. Several investigations using \bifrost{} have been done or are under way, several of these including PhD students who have been able to use this `state of the art' numerical code with relatively little instruction. There are a number of modules being developed, including a module that can follow the ionization states of heavy elements and a module that introduces a modified Ohm's law. There is a trend towards computer systems using the large raw floating point performance of Grapical Processing Units (GPUs), and a new parallelization module for such systems is also under development. 

The very good scaling performance of \bifrost{} makes it possible to make simulations of stellar atmospheres with a very large resolution while still encompassing a large enough volume to make  the simulation realistic, and include a large number of physical effects. The results can be used to predict observational effects, which might earlier have lead to wrong diagnostics of the physical parameters in the solar atmosphere. 
 
\begin{acknowledgements}
We would like to thank the referee for very useful comments,
  that helped us make improvements. We would like to thank Robert F. Stein, Andrew McMurry and Colin Rosenthal for working out the formalism for the characteristic boundary conditions.
This research was supported by the Research Council of Norway through
the grant ``Solar Atmospheric Modelling'' and 
through grants of computing time from the Programme for Supercomputing
and through grants SMD-07-0434, SMD-08-0743, SMD-09-1128, SMD-09-1336,
and SMD-10-1622 from the High End Computing (HEC) division of NASA.
The authors wish to thank PRACE for opportunity to run experiments on
HPC centers in Europe. J.L. acknowledges financial support from the European Commission
through the SOLAIRE Network (MTRN-CT-2006-035484) and  from the Netherlands Organization for
  Scientific Research (NWO).
\end{acknowledgements}

\bibliographystyle{aa}
\bibliography{manuscript}

\appendix

% This appendix should span both columns, because of the long equations

\section{Characteristic boundary conditions}\label{app:char_bound}
The boundary equations in terms of the primitive variables
$(\rho,\mathbf{u},e,\mathbf{B})$ can be written in the following form

\begin{eqnarray}
\ddt{\rho} &=& -{1\over c_\mathrm{s}^2}\left[{1\over\rho}\dPde
                 d_2+{\rho\alpha_+\over 2}(d_5+d_6)
                    +{\rho\alpha_-\over 2}(d_7+d_8)\right] \nonumber \\
          &   &  \qquad -(\uH\cdot\dH)\rho-\rho\dH\cdot\uH
\end{eqnarray}
\begin{eqnarray}
\ddt{u_x} & = &  -{s_z\over 2}\left[R_y(-d_3+d_4)
                                  +{c_-\alpha_-\over
                                    c_\mathrm{s}^2}R_x(-d_5+d_6) \right.\nonumber\\
                                  &&\qquad\qquad\left.+{c_+\alpha_+\over 
                                    c_\mathrm{s}^2}R_x(d_7-d_8)\right]
                                \nonumber \\
          &   &  -{1\over\rho}\dd{\left(P+{B^2\over
                       2\mu_0}\right)}{x}-(\uH\cdot\dH)u_x
                 +\om{1}(\BH\cdot\dH)B_x \\
\ddt{u_y} & = &  -{s_z\over 2}\left[R_x(d_3-d_4)
                                  +{c_-\alpha_-\over c_\mathrm{s}^2}R_y(-d_5+d_6)\right.\nonumber\\
                                  &&\qquad\qquad\left.+{c_+\alpha_+\over
                                    c_\mathrm{s}^2}R_y(d_7-d_8)\right]
                                \nonumber \\
          &   &  -{1\over\rho}\dd{\left(P+{B^2\over
                       2\mu_0}\right)}{y}-(\uH\cdot\dH)u_y
                 +\om{1}(\BH\cdot\dH)B_y
\end{eqnarray}
\begin{eqnarray}
\ddt{u_z} & = & -{1\over 2c_\mathrm{s}^2}
    \left[c_+\alpha_+(d_5-d_6)+c_-\alpha_-(d_7+d_8)\right]
                       \nonumber \\
          &   & \qquad -(\uH\cdot\dH)u_z+g+\om{1}(\BH\cdot\dH)B_z \\
\end{eqnarray}
\begin{eqnarray}
\ddt{e} & = & -{1\over c_\mathrm{s}^2}\left[{1\over\rho}\dPdr
                            +\half(e+P)\alpha_+(d5+\alpha_+d6)\right.\nonumber\\
                         &&\left.\qquad +\half(e+P)\alpha_-(d_7+d_8)\right] \nonumber \\
        &  & \quad -\dH\cdot(e\uH)-P\dH\cdot\uH+Q \\
\ddt{B_x} & = & - {\sm\over 2}\left[R_y(d_3+d_4)
                           +{\alpha_-\over c_\mathrm{s}}R_x(d_5+d_6)\right.\nonumber\\
                          &&\left.\qquad-{\alpha_+\over c_\mathrm{s}}R_x(d_7+d_8)\right] \nonumber
                           \\
        &  & \quad -(\uH\cdot\dH)B_x+B_x(\dH\cdot\uH)+(\BH\cdot\dH)u_x \\
\ddt{B_y} & = & - {\sm\over 2}\left[-R_x(d_3+d_4)
                           +{\alpha_-\over c_\mathrm{s}}R_y(d_5+d_6)\right.\nonumber\\
                           &&\left.\qquad-{\alpha_+\over c_\mathrm{s}}R_y(d_7+d_8)\right] \nonumber
                           \\
        &  & \quad -(\uH\cdot\dH)B_y+B_y(\dH\cdot\uH)+(\BH\cdot\dH)u_y \\
\ddt{B_z} & = & -(\uH\cdot\dH)B_z-B_z(\dH\cdot\uH)+(\BH\cdot\dH)u_z 
\end{eqnarray}
where we have defined the quantities
\begin{eqnarray*}
s_z & = & \signBz \\
R_x & = & \frac{B_x}{B_h} \qquad
R_y  =  \frac{B_y}{B_h} \\
\alpha_+^2 & = & \frac{c_\mathrm{s}^2 - c_-^2}{c_+^2 - c_-^2} \qquad
\alpha_-^2  =  \frac{c_+^2 - c_\mathrm{s}^2}{c_+^2 - c_-^2} \\
\end{eqnarray*}
using the velocities
\begin{eqnarray*}
c_\mathrm{s}^2 & = & \dPdr + \frac{e + P}{\rho} \dPde \\
c_\mathrm{a}^2 & = & \om{B^2} \\
c_z^2 & = & \om{B_z^2} \\
c_+^2 & = & \frac{c_\mathrm{a}^2 + c_\mathrm{s}^2}{2} + \sqrt{\left(\frac{c_\mathrm{a}^2 +
c_\mathrm{s}^2}{2}\right)^2 - c_z^2 c_\mathrm{s}^2} \\
c_-^2 & = & \frac{c_\mathrm{a}^2 + c_\mathrm{s}^2}{2} - \sqrt{\left(\frac{c_\mathrm{a}^2 +
c_\mathrm{s}^2}{2}\right)^2 - c_z^2 c_\mathrm{s}^2}.
\end{eqnarray*}
Note the unorthodox ordering of the characteristics: by convention
these are usually ordered by the amplitude of the characteristic
speeds $\lambda_i$, this was not known by us at the time these equations were derived.
Therefore, the characteristic $z$ derivatives $\mathbf{d}$ are given by 
\begin{eqnarray}
d_1 & = & u_z \ddz{B_z} \\
d_2 & = & u_z (e+P)\ddz{\rho}-\rho\ddz{e} \\
d_3 & = & (u_z+c_z) \left(-s_zR_y\ddz{u_x}+s_zR_x\ddz{u_y} \right.\\
         &&\left.\qquad+{R_y\over\sm}\ddz{B_x}-{R_x\over\sm}\ddz{B_y}\right) \\
d_4 & = & (u_z-c_z) \left(s_zR_y\ddz{u_x}-s_zR_x\ddz{u_y}\right.\\
          &&\left.\qquad+{R_y\over\sm}\ddz{B_x}-{R_x\over\sm}\ddz{B_y}\right) \\
d_5 & = & (u_z+c_+) \left({\alpha_+\over\rho}\ddz{P}
           -s_zR_xc_-\alpha_-\ddz{u_x}-s_zR_yc_-\alpha_-\ddz{u_y}\right.\nonumber\\
           &   &  \left.+c_+\alpha_+\ddz{u_z}
    +\som{R_xc_\mathrm{s}\alpha_-}\ddz{B_x}+\som{R_yc_\mathrm{s}\alpha_-}\ddz{B_y}\right) \\
d_6 & = & (u_z-c_+) \left({\alpha_+\over\rho}\ddz{P}
           +s_zR_xc_-\alpha_-\ddz{u_x}+s_zR_yc_-\alpha_-\ddz{u_y}\right.\nonumber\\
           & &\left.-c_+\alpha_+\ddz{u_z}
       +\som{R_xc_\mathrm{s}\alpha_-}\ddz{B_x}+\som{R_yc_\mathrm{s}\alpha_-}\ddz{B_y}\right) \\
d_7 & = & (u_z+c_-) \left({\alpha_-\over\rho}\ddz{P}
           +s_zR_xc_+\alpha_+\ddz{u_x}+s_zR_yc_+\alpha_+\ddz{u_y}\right. \nonumber\\
           &   &  \left.+c_-\alpha_-\ddz{u_z}
    -\som{R_xc_\mathrm{s}\alpha_+}\ddz{B_x}-\som{R_yc_\mathrm{s}\alpha_+}\ddz{B_y}\right) \\
d_8 & = & (u_z-c_-) \left({\alpha_-\over\rho}\ddz{P}
           -s_zR_xc_+\alpha_+\ddz{u_x}-s_zR_yc_+\alpha_+\ddz{u_y}\right. \nonumber\\
           &   &  \left.-c_-\alpha_-\ddz{u_z}
    -\som{R_xc_\mathrm{s}\alpha_+}\ddz{B_x}-\som{R_yc_\mathrm{s}\alpha_+}\ddz{B_y}\right)\textrm{\ .}
\end{eqnarray}

Along inflowing characteristics the characteristic derivatives
$\mathbf{d}$ 
are changed to provide
transmitting boundaries by requiring that the incoming characteristics
remaining constant. Thus, we require that the boundary conditions for
the incoming characteristics must satisfy the static case where $\uu =
0$ and $\ddt{\UU} = 0$, so that
\begin{equation}\label{eq:di}
d_i = (\SM^{-1} \CC^{(\uu=0)})_i\textrm{\ .}
\end{equation}
The components of $\CC^{(\uu=0)}$ are $0$ for the magnetic field equation and the density
equation.  For the other equations:
\begin{eqnarray*}
C_e^{(\uu=0)} & = & Q \label{eq:Ceu0}\\
C_{u_x}^{(\uu=0)} & = & -\frac{1}{\rho}\dd{}{x}\left(P + \frac{B^2}{2\mu_0}\right) + \om{1}(\Bh \cdot \delh) B_x \\
C_{u_y}^{(\uu=0)} & = & -\frac{1}{\rho}\dd{}{y}\left(P + \frac{B^2}{2\mu_0}\right) + \om{1}(\Bh \cdot \delh) B_y \\
C_{u_z}^{(\uu=0)} & = & g + \om{1}(\Bh \cdot \delh) B_z \label{eq:Cuzu0}\textrm{\ .}
\end{eqnarray*}
Assuming $Q = 0$, the non-zero incoming characteristic derivative
vector $\mathbf{d}$ can be calculated to be:
\begin{eqnarray}
d^i_3 & = & \frac{s_z}{\rho}\left[
-(\Rh\times\delh)\left(P + \frac{B_z^2}{2\mu_0}\right)
+\frac{B_H}{\mu_0}(\delh\times\Bh)\right] \\
d^i_4 & = & -d^i_3 \\
d^i_5 & = & c_+ \alpha_+ \left(g + \om{1}(\Bh \cdot \delh)
  B_z\right) \\  && \quad +
c_- \alpha_- \frac{s_z}{\rho}(\Rh\cdot\delh)\left(P + \frac{B_z^2}{2\mu_0}\right) \\
d^i_6 & = & -d^i_5 \\
d^i_7 & = & c_- \alpha_- \left(g + \om{1}(\Bh \cdot \delh) B_z\right)
\\ &&\quad -
c_+ \alpha_+ \frac{s_z}{\rho}(\Rh\cdot\delh)\left(P + \frac{B_z^2}{2\mu_0}\right) \\
d^i_8 & = & -d^i_7\textrm{\ .}
\end{eqnarray}
Then either $d_i$ or $d^i_i$ is chosen depending on the sign of
$\lambda_i$ at the boundary; {\it i.e.} whether or not the
characteristic is in- or out-flowing.

\section{Time-dependent \htwo\ formation}\label{app:hion}

The module that computes non-equilibrium hydrogen ionization has been
extended to include a ninth equation for time-dependent \htwo\
formation. It is given by:
\begin{equation} 
F_9 = \frac{\nhtwo}{\nhtwoo} - \frac{\Delta t}{\nhtwoo}
 \left( \Cthreeh \none^3 - \Chtwoh \nhtwo \none \right) -1 = 0\textrm{\ ,}
\end{equation}
with $\nhtwoo$\ the \htwo\ population of the previous timestep, $\Delta
t$ the timestep and $\none$\ the population of atomic hydrogen in the
ground state. The rate coefficients are given by
\begin{eqnarray} 
\Chtwoh &=& \alpha \left( \frac{T}{300 \,\mathrm{K}} \right)^\beta \mathrm{e}^{-\gamma/T}\textrm{\ ,} \\
\Cthreeh  &=& \frac{\Chtwoh}{K(T)} \textrm{\ .}
\end{eqnarray}
The values for $\alpha$, $\beta$ and $\gamma$ are taken from the UMIST
database
\citep[\url{www.udfa.net}]{2007A&A...466.1197W}; %UMIST 2006
$K(T)$ is the chemical equilibrium constant taken from
\citet{1973A&A....23..411T}.
The derivatives of the functional $F_9$ with respect to the dependent
variables are
\begin{eqnarray} 
\dd{F_9}{T} &=& \frac{\Delta t}{\nhtwoo}
 \left( \dd{\Cthreeh}{T} \none^3 - \dd{\Chtwoh}{T} \nhtwo \none \right)\textrm{\ ,}\\
\dd{F_9}{\none}  &=& - \frac{\Delta t}{\nhtwoo}
 \left( 3 \Cthreeh \none^2 - \Chtwoh \nhtwo \right) \textrm{\ ,}\\
\dd{F_9}{\nhtwo}  &=& \frac{1}{\nhtwoo} - \frac{\Delta t}{\nhtwoo} \Chtwoh \none\textrm{\ .}
\end{eqnarray}

The rate equation for $\none$ from 
\citet{Leenaarts+etal07}
is modified to include source and sink terms due to \htwo:
\begin{equation}
F_3 = \frac{\none}{\noneo}  - \frac{\Delta t}{\noneo} 
   \left( \sum_{j=1}^6 n_j P_{j1} +  \Chtwoh \nhtwo \none -  \Cthreeh
   \none^3 \right) -1 = 0\textrm{\ ,}
\end{equation}
with $\noneo$ the ground state hydrogen population from the previous timestep.
The derivatives of this equation and equations expressing energy
conservation and hydrogen nucleus conservation
\citep[see][]{Leenaarts+etal07}
 are modified
correspondingly.

\end{document}